&pdflatex
\documentclass[3p]{elsarticle}

\usepackage{url}
\usepackage{amssymb}
\usepackage{amsmath}
\usepackage{amsthm}
\usepackage{nicefrac}
\usepackage{tabu}
\usepackage{caption}
\usepackage{algpseudocode}
\usepackage{algorithm}
\usepackage{tabto}

\usepackage{color}

\begin{document}\sloppy

\let\originalleft\left
\let\originalright\right
\renewcommand{\left}{\mathopen{}\mathclose\bgroup\originalleft}
\renewcommand{\right}{\aftergroup\egroup\originalright}

\newcommand{\cpp}{{\nolinebreak C\texttt{++} }}

\newcommand{\BigO}[1]{\mathop{}\!O{\left(#1\right)}}

\newcommand{\Del}[1]{\operatorname{Del}\left(#1\right)}
\newcommand{\Vor}[1]{\operatorname{Vor}\left(#1\right)}

\newcommand{\reviewerA}{}
\newcommand{\reviewerB}{}
\newcommand{\reviewerC}{}

\theoremstyle{definition}

\newtheorem{remark}{Remark}
\newtheorem{definition}{Definition}

\begin{frontmatter}

\title{Generalised primal-dual grids for unstructured co-volume schemes}

\author[a,b]{Darren Engwirda\corref{cor1}}
\ead{darren.engwirda@columbia.edu}

\address[a]{Center for Climate Systems Research, Columbia University, New York City, NY 10025, USA}

\address[b]{NASA Goddard Institute for Space Studies, 2880 Broadway, New York City, NY 10025, USA}

\begin{abstract}
The generation of high-quality staggered unstructured grids is considered, leading to the development of a new optimisation-based strategy designed to construct weighted `Regular-Power' tessellations appropriate for co-volume type numerical discretisation techniques. This new framework aims to extend the conventional Delaunay-Voronoi primal-dual structure; seeking to assemble generalised orthogonal tessellations with enhanced geometric quality. The construction of these grids is motivated by the desire to improve the performance and accuracy of numerical methods based on unstructured co-volume type schemes, including various staggered grid techniques for the simulation of fluid dynamics and hyperbolic transport. In this study, a new hybrid optimisation strategy is proposed; seeking to optimise the geometry, topology and weights associated with general, two-dimensional Regular-Power tessellations using a combination of gradient-ascent and energy-based techniques. The performance of this new method is tested experimentally, with a range of complex, multi-resolution primal-dual grids generated for various coastal and regional ocean modelling applications.

\end{abstract}

\cortext[cor1]{Corresponding author. Tel.: +1-212-678-5521}
\end{frontmatter}

\section{Introduction}
\label{section_introduction}

Co-volume techniques, in which systems of partial differential equations are discretised using a pair of staggered, orthogonal computational grids, represent an important class of numerical methods; leading to a range of schemes for the simulation of complex physical phenomena, including various problems in fluid dynamics and hyperbolic transport. Such schemes include the classical Marker And Cell (MAC) method \cite{harlow1965numerical} for the solution of the Navier-Stokes equations, and various Arakawa-type schemes \cite{arakawa1977computational} for the solution of the primitive equations in geophysical fluid dynamics. Due to the underlying staggered discretisation strategy, such schemes are typically imbued with a range of desirable conservation properties, preserving, for example, various mass-, energy- and rotation-based physical invariants in the simulation of fluid phenomena. Such behaviour makes co-volume schemes natural choices for the simulation of geophysical flows, leading to the development of high-fidelity algorithms for numerical weather prediction, ocean modelling, and planetary climate dynamics.

In an unstructured setting, the development of co-volume schemes leads to a challenging grid generation problem, requiring the construction of a compatible \textit{primal-dual pair} $(\mathcal{T},\mathcal{D})$ that tessellates a given domain $\Omega \subset \mathbb{R}^{d}$ into a mesh of discrete elements. The pair $(\mathcal{T},\mathcal{D})$ is required to be a dual structure; a tessellation of the discrete points $\mathbf{x} \in \mathbb{R}^{d}$ into a primal \textit{simplicial triangulation} $\mathcal{T}$ and its dual \textit{polyhedral complex} $\mathcal{D}$. Here, duality is expressed as {\reviewerA a relationship} between the faces of the triangulation and its dual, such that each $d$-simplex is associated with a dual vertex (a $0$-cell), each $(d-1)$-simplex is associated with a dual edge (a $1$-cell) spanning between the two dual vertices associated with the adjacent $d$-simplexes, and so on, as per \cite{memari2011parametrization}. Such structures are also typically required to satisfy a number of additional conditions: preserving \textit{orthogonality} between adjacent faces in $\mathcal{T}$ and $\mathcal{D}$, conforming to general, user-defined  constraints on \textit{grid-spacing}, and consisting of \textit{optimally-shaped} grid-cells, designed to minimise the error associated with discrete numerical operators. See Figure~\ref{figure_primal_dual_TRSK} for additional detail.

Despite the maturity of approaches for the generation of high-quality unstructured triangulations \cite{shewchuk1996triangle,si2015tetgen,jamin2015cgalmesh,schoberl1997netgen,geuzaine2009gmsh,engwirda2017jigsaw}, the construction of primal-dual structures for co-volume techniques remains a challenging task, with the quality of the dual polyhedral grid typically lagging behind that of the primal triangulation. In this study, a new framework for the generation of high-quality primal-dual grids is developed, based on a class of weighted Voronoi-type tessellations known as \textit{power diagrams} \cite{aurenhammer1987power,edelsbrunner2012algorithms}. While this new approach is applicable to co-volume type schemes in general, a particular motivation is the generation of optimal, multi-resolution grids for unstructured ocean modelling and climate dynamics, focusing on techniques appropriate for the mimetic finite-difference/volume schemes presented by, for example, Ringler et al \cite{ringler2010unified,ringler2013multi}, Thurburn et al \cite{thurburn2009}, and Korn et al \cite{KORN2017525,KORN2017156}. In the present study, a two-dimensional implementation is developed, leading to the construction of complex, multi-resolution grids for coastal and regional ocean modelling applications.

\subsection{Constraints on co-volume grids}
\label{section_grid_constraints}

\medskip


\begin{figure*}
  \centering
  \includegraphics[width=.9125\textwidth]{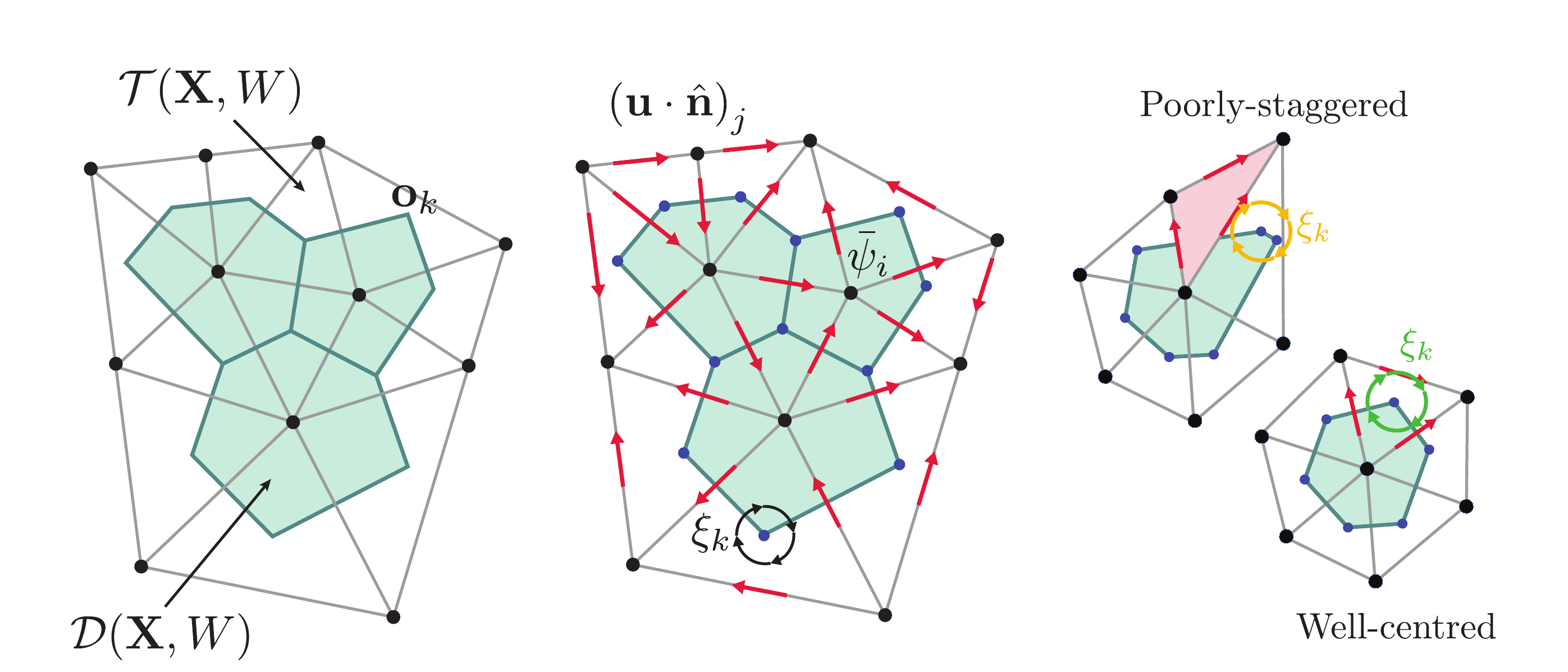}
  
  \caption{Anatomy of an unstructured co-volume scheme for geophysical fluid dynamics, illustrating: (a) locally-orthogonal primal-dual grid staggering, (b) the mimetic finite-difference/volume formulation of Ringler et al \cite{ringler2010unified,ringler2013multi}, and (c) a comparison of \textit{well-centred} and \textit{poorly-staggered} configurations. In (a), polygonal cells in the dual grid are formed by joining the orthocentres $\mathbf{o}_{k}$ associated with adjacent primal triangles. In (b), the unstructured TRSK scheme employs conservative \textit{cell-centred} tracer quantities $\bar{\psi}_{i}$, \textit{edge-centred} normal velocity components $(\mathbf{u}\cdot\hat{\mathbf{n}})_{j}$, and auxiliary \textit{vertex-centred} vorticity variables $\xi_{k}$. In (c), the \textit{well-centred} (lower) and \textit{poorly-staggered} (upper) configurations differ in the relationship between dual vertices and primal triangles. In the well-centred configuration, all dual vertices are \textit{interior} to their parent triangles. In the poorly-staggered configuration, a dual vertex is exterior to its associated triangle (shaded). Note that adjacent primal/dual edges do not intersect in the poorly-staggered configuration.
  }
  \label{figure_primal_dual_TRSK}
\end{figure*}

Due to the nature of the underlying dual discretisation strategy, co-volume schemes impose a unique set of requirements on the associated meshing problem; requiring high-quality primal-dual tessellations that satisfy a number of geometrical constraints. While spatial-staggering and pair-wise orthogonality are basic prerequisites for such schemes, additional conditions are motivated by a desire to maximise the accuracy, stability and efficiency of the associated co-volume formulation. Here, such considerations are explored in the context of the mimetic finite-difference/volume scheme presented by Ringler et al \cite{ringler2010unified,ringler2013multi} and Thurburn et al \cite{thurburn2009} for the simulation of geophysical flows, though similar arguments are applicable to the optimisation of co-volume formulations in general.

The `TRSK' formulation of Ringler et al is a generalisation of the structured Arakawa C-grid scheme to unstructured primal-dual pairs. In this scheme, fluid pressure, depth, and density degrees of freedom are located within the dual polygonal control volumes, and a set of orthogonal velocity vectors are positioned along the primal grid edges. Additional vorticity degrees of freedom are placed at the vertices of the dual grid cells. Such an arrangement facilitates the construction of a standard conservative finite-volume type scheme for the transport of cell-centred fluid properties, and a mimetic finite-difference formulation for the evolution of velocity components. Vorticity is computed by considering the discrete circulation of velocity about primal triangles. See Figure~\ref{figure_primal_dual_TRSK} for details. Overall, this scheme is known to posses a variety of desirable conservation properties, conserving mass, potential vorticity and enstrophy, and preserving exact geostrophic balance \cite{ringler2010unified}. The TRSK scheme is currently employed in the Model for Prediction Across Scales (MPAS) for both atmospheric and oceanic modelling \cite{skamarock2012multiscale,ringler2010unified,ringler2013multi}.

Despite its unstructured nature, such a formulation is not applicable to general unstructured tessellations; requiring that a number of auxiliary geometrical constraints also be satisfied. Specifically, such schemes necessitate the use of grids that are not only staggered and pair-wise orthogonal but also \textit{well-centred} and \textit{centroidal}. These additional conditions are constraints on the geometry of the underlying primal triangles and their dual polygons. 

A well-centred grid \cite{vanderzee2008well,vanderzee2010well} is one in which all dual vertices lie within the interior of their associated primal triangles. Such a condition guarantees that adjacent primal and dual edges intersect; inducing a dual structure that is \textit{nicely staggered}. In the context of the unstructured TRSK scheme described previously, this configuration guarantees that a consistent computational stencil exists for evaluation of the various discrete numerical operators present in the governing equations, allowing such schemes to preserve physical invariants associated with the flow. Centroidal tessellations are those in which primal and dual vertices are positioned at the centroids of the associated staggered grid cells, with the vertices of primal triangles located at the centres-of-mass of the associated dual polyhedra and visa versa. Ideally, \textit{centroidal} staggering should also be exhibited by the lower dimensional faces of the pair $(\mathcal{T},\mathcal{D})$, with adjacent faces intersecting at their local centroids. Such structures can be used to construct compact spatial discretisation schemes that exhibit optimal numerical performance. The unstructured TRSK scheme described previously is known to achieve second-order accurate convergence on uniform primal-dual grids that are well-centred, centroidal and pair-wise orthogonal \cite{ringler2010unified}.

\subsection{Related work}

\medskip

A number of approaches for the generation of high-quality primal-dual grids have previously been pursued, typically focusing on the optimisation of conventional Delaunay-Voronoi type pairs. Such methods include a range of well-known \textit{variational} techniques, including methods based on Centroidal Voronoi Tessellation (CVT) \cite{du1999centroidal,du2002grid} and Optimal Delaunay Triangulation (ODT) \cite{alliez2005variational,chen2004optimal,chen2011efficient} that seek to reposition primal vertices to minimise {\reviewerA an \textit{objective function} that depends on the geometry and topology of the grid}. While sophisticated CVT/ODT schemes have been shown to generate high quality centroidal Voronoi-type tessellations, the construction of structures that are additionally well-centred has proven to be a difficult task; requiring the generation of so-called \textit{non-obtuse} triangulations in which all angles in the primal mesh are bounded below $90^\circ$. In \cite{vanderzee2008well,vanderzee2010well}, Vanderzee et al have shown that a non-linear optimisation-based strategy can be used to create such non-obtuse triangulations, though the resulting primal-dual tessellations do not appear to be centroidal or necessarily of optimal quality.

More recently, \textit{generalised} orthogonal primal-dual pairs have been investigated \cite{memari2011parametrization}, with Mullen et al developing the so-called Hodge-Optimized Triangulation (HOT) \cite{mullen2011hot,goes2014weighted} framework in the context of computer graphics applications. The HOT formulation is a generalised variational approach, in which the minimisation of a coupled primal-dual energy functional is undertaken. Such a formulation is designed to optimise the performance of an associated Discrete Exterior Calculus (DEC) type formulation; aiming to generate generalised primal-dual pairs that are typically centroidal, well-centred and pair-wise orthogonal. In this work, Mullen et al employ a weighted primal-dual structure based on the power diagram and its associated regular triangulation, and construct a coupled, optimisation-based approach designed to minimise the combined HOT functional. 

{\reviewerC
The use of weighted primal-dual pairs has also been pursued by Walton et al \cite{walton2017advances,walton2013a,walton2013b} in which a multi-paradigm optimisation scheme is used to improve the well-centredness of two- and three-dimensional unstructured grids for co-volume type aerospace modelling. In this work, a novel modified `cuckoo' type optimisation strategy is employed to minimise a dual mesh quality metric that measures the offset between dual vertices and their associated primal element centroids. These operations are combined with topological modifications, in which clusters of primal elements associated with sufficiently small dual edges are merged into polyhedra. Walton et al report good performance for a variety of two- and three-dimensional problems, and demonstrate that optimised primal-dual meshes often improve the stability and efficiency of co-volume type solvers for problems in aerodynamics and electromagnetism.
}  

In the context of unstructured ocean-modelling, the generation of optimal co-volume grids have typically focused on the application of CVT-based methods, with Jacobsen et al presenting the massively-parallel MPI-SCVT algorithm \cite{jacobsen2013parallel} for the generation of graded, Voronoi-type meshes on the sphere. These grids have been utilised for various multi-resolution studies using the Model for Prediction Across Scales (MPAS-O) \cite{ringler2008multiresolution,ringler2013multi}. Despite producing high-quality centroidal Voronoi structures, the CVT-based approach has not been found to reliably generate well-centred grids; leading to issues with the preservation of rotational invariants in the simulation of geophysical flows, as discussed previously. An alternative, hybrid approach has recently been described by the author in \cite{engwirda2017jigsaw}, in which a Frontal-Delaunay refinement procedure is used to provide a high-quality initial Delaunay-Voronoi pair that is subsequently improved through an optimisation phase. Such an approach has been shown to successfully generate a range of complex, centroidal and well-centred Voronoi-type grids on the sphere.

In the present study, a generalisation of the approach presented by the author in \cite{engwirda2017jigsaw} is pursued, but in which a weighted primal-dual structure is optimised in place of the conventional Delaunay-Voronoi pair. {\reviewerB Firstly, a new dual mesh quality metric is introduced, designed to minimise the discretisation errors associated with a staggered mimetic discretisation of the primitive equations on unstructured grids. Specifically, the numerical errors associated with the discrete $\nabla\cdot(\cdot)$, $\nabla\times(\cdot)$ and $\nabla(\cdot)$ operators are analysed, and are used to inform the construction of a new dual grid quality metric. Mesh generation is cast as a coupled optimisation problem, seeking to select a set of vertex positions, weights and an associated mesh topology that maximises a set of primal-dual quality metrics. While the approach presented here is related to both the HOT framework of Mullen et al \cite{mullen2011hot,goes2014weighted} and the optimisation-based strategies of Walton et al \cite{walton2017advances,walton2013a,walton2013b}, a number of important differences exist. Here, a \textit{monotone} optimisation scheme is developed, ensuring that mesh quality metrics are increased monotonically throughout the optimisation process, guaranteeing that no new low-quality elements are created. Secondly, through the construction of a new \textit{scale-invariant} dual quality metric, the treatment of non-uniform meshes is realised. A coupled optimisation schedule is employed here, interleaving various edge refinement and collapse operations, topological transformations and updates to grid geometry and weights. The performance of the new scheme is analysed experimentally, demonstrating that significantly enhanced primal-dual complexes can be generated for a range of challenging problems involving complex boundary definitions and non-uniform mesh-spacing constraints.} 


\section{Regular triangulations \& power diagrams}
\label{section_power_dual}

\begin{figure*}
  \centering
  \includegraphics[width=.9125\textwidth]{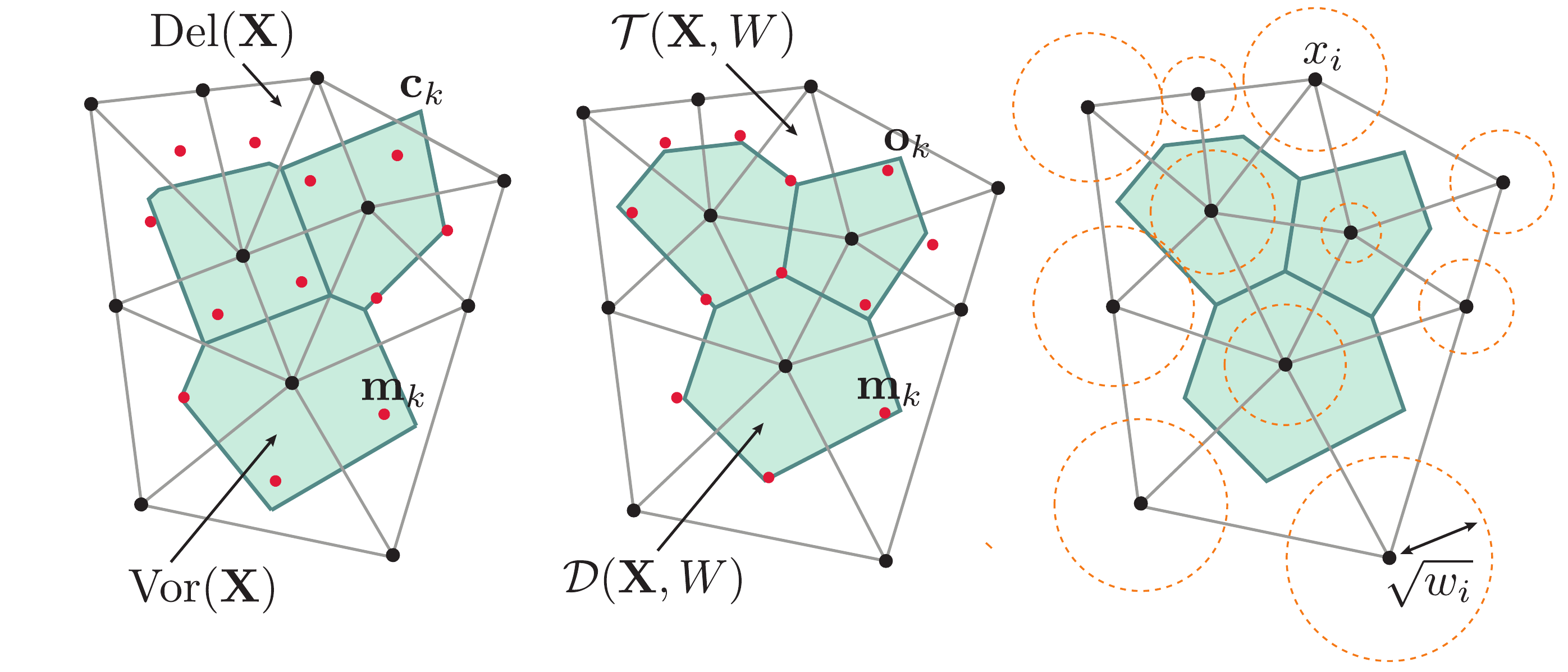}
  
  \caption{A comparison of various unstructured primal-dual pairs, showing: (a) a standard Delaunay-Voronoi tessellation, (b) an optimised Regular-Power structure, and (c) the distribution of vertex weights employed in (b). The position of triangle centroids and poorly-staggered elements is highlighted in (a) and (b), demonstrating that the weighted Regular-Power grid is both more centroidal and better self-centred than the corresponding Delaunay-Voronoi mesh. Note that though the underlying Delaunay triangulation is of high quality and is approximately centroidal, its corresponding Voronoi dual incorporates a range of undesirable features, including short edge-segments and non-optimal staggering.  
  }
  \label{figure_primal_dual_2}
\end{figure*}

Before describing the new primal-dual grid generation algorithm in full, several concepts associated with the construction of \textit{regular triangulations} and their associated dual \textit{power diagrams} are presented. These structures can be viewed as a weighted generalisation of the conventional Delaunay tessellation and its dual Voronoi complex. {\reviewerA Such structures are also sometimes referred to as \textit{Laguerre diagrams} or \textit{Dirichlet cell complexes} (see, for example \cite{aurenhammer1987power})}. 

\begin{definition}[Weighted points \& power distance]
A \textit{weighted} point set is defined as a pair $(\mathbf{X},W) = \{(\mathbf{x}_{1},w_{1}),(\mathbf{x}_{2},w_{2}),\dots,(\mathbf{x}_{n},w_{n})\}$, where {\reviewerB$\{\mathbf{x}_{i}\} \subset \mathbb{R}^{d}$} are a set of points embedded in $d$-dimensional Euclidean space and {\reviewerB$\{w_{i}\} \subset \mathbb{R}$} are an associated set of scalar weights. {\reviewerB The \textit{power distance} \cite{aurenhammer1987power}}, herein denoted $\pi_{i}(\mathbf{x})$, between an unweighted point $\mathbf{x} \in \mathbb{R}^{d}$ and a weighted point $(\mathbf{x}_{i},w_{i})$ is defined as $\pi_{i}(\mathbf{x}) = \|\mathbf{x}-\mathbf{x}_{i}\|^{2} - w_{i}$, where $\|\cdot\|$ is the standard Euclidean distance operator.
\end{definition}

\begin{definition}[Regular-Power tessellation]
Given a weighted point set $(\mathbf{X},W)$, the {\reviewerB \textit{power complex} \cite{aurenhammer1987power}} $\mathcal{D}(\mathbf{X},W)$ is the union of polyhedral cells $\{D_{i}\}$ , where each $D_{i} = \left\{\mathbf{x} \in \mathbb{R}^{d} \,\, | \,\, \pi_{i}(\mathbf{x}) < \pi_{j}(\mathbf{x}) \,, \,\forall\, j \neq i \right\}$. The associated primal complex $\mathcal{T}(\mathbf{X},W)$ is a simplicial triangulation of the weighted points $(\mathbf{X},W)$, consisting of the union of simplexes $\{\tau_{k}\}$, {\reviewerB where each $\tau_{k}$ contains the vertices $\{\mathbf{x}_{1},\mathbf{x}_{2},\dots,\mathbf{x}_{k}\} \in \mathbf{X}$ iff $\bigcap_{\,j=1}^{\,j=k} D_{j} \neq \emptyset$ \cite{mullen2011hot}}. The primal complex is known as a \textit{regular triangulation} of the weighted points $(\mathbf{X},W)$.
\end{definition}

The power complex is a Voronoi-like subdivision of space, where each cell $D_{i} \in \mathcal{D}$ defines a convex region $\mathbf{x} \subseteq \mathbb{R}^{d}$ for which the weighted point $(\mathbf{x}_{i},w_{i})$ is \textit{closer}, in a weighted sense, than all other points in $(\mathbf{X},W)$. Such structures, along with their dual regular triangulations, offer a generalised framework for the construction of meshes for co-volume schemes \cite{memari2011parametrization,mullen2011hot,goes2014weighted}; providing a family of compatible, pair-wise orthogonal primal-dual structures paramaterised by the set of scalar weights $W$. In this work, the selection of an \textit{optimal} set of weights is sought; tuning the geometry of the power cells to construct a primal-dual structure with optimal characteristics. See Figure~\ref{figure_primal_dual_2} for details. Detailed analysis of the properties of power complexes and regular triangulations has been investigated previously, and the reader is referred to, for example, studies by Aurenhammer \cite{aurenhammer1987power} and Edelsbrunner \cite{edelsbrunner2012algorithms} for additional detail. 

Given a weighted point set $(\mathbf{X},W)$, the associated Regular-Power tessellation is fully specified by determination of the position of vertices associated with the polyhedral dual. These points are known as \textit{orthocentres}, and can be viewed as a weighted generalisation of the circumcentres associated with the standard Delaunay-Voronoi structure. In the present work, orthocentres are constructed with respect to both the edges and faces of the primal triangulation, as detailed in the following sections.

\subsection{Weighted `edge' orthocentres}

\medskip

Given a triangle $\tau_{i}$ in the weighted primal mesh $\mathcal{T}(\mathbf{X},W)$, the \textit{orthocentre} associated with the edge $\{(\mathbf{x}_{i},w_{i}),(\mathbf{x}_{j},w_{j})\} \in \tau_{i}$ is the point $\mathbf{o}_{e}$ of equal power distance to the endpoints
\begin{equation}
\label{eqn_orthocentre_edge_1}
\|\mathbf{x}_{i}-\mathbf{o}_{e}\|^{2} - w_{i} =
\|\mathbf{x}_{j}-\mathbf{o}_{e}\|^{2} - w_{j}\,.
\end{equation}
Constraining the point $\mathbf{o}_{e}$ to the edge vector, and solving for the associated line parameter $t\in\mathbb{R}$ leads to an expression for $\mathbf{o}_{e}$, such that
\begin{gather}
\label{eqn_orthocentre_edge_2}
\mathbf{o}_{e} = \mathbf{x}_{i} + t\,(\mathbf{x}_{j}-\mathbf{x}_{i})\,, \quad 0 \leq t \leq 1\,,
\\[1.33ex]
t = \frac{1}{2} 
\left(\frac{w_{i}-w_{j}+\|\mathbf{x}_{i}-\mathbf{x}_{j}\|^{2}}{\|\mathbf{x}_{i}-\mathbf{x}_{j}\|^{2}}\right)\,.
\end{gather}
Note that in unweighted configurations with $w_{i} = w_{j} = 0$, the expression for $\mathbf{o}_{e}$ reduces to a simple equation for the edge midpoint $\mathbf{m}_{e}$, with $t = \frac{1}{2}$, as expected. Note also that $\mathbf{o}_{e}$ is sensitive to the \textit{difference} in the weights $w_{i}$ and $w_{j}$ rather than their magnitude, with equally weighted configurations $w_{i} = w_{j} = \beta \in \mathbb{R}$ again recovering $\mathbf{o}_{e} = \mathbf{m}_{e}$.

\subsection{Weighted `face' orthocentres}

\medskip

Given a triangle $\tau_{i}$ in the weighted primal mesh $\mathcal{T}(\mathbf{X},W)$, the \textit{orthocentre} associated with the face $\{(\mathbf{x}_{i},w_{i}),(\mathbf{x}_{j},w_{j}),(\mathbf{x}_{k},w_{k})\} \in \tau_{i}$ is the point $\mathbf{o}_{f}$ of equal power distance to the three corner vertices
\begin{equation}
\label{eqn_orthocentre_face_1}
\|\mathbf{x}_{i}-\mathbf{o}_{f}\|^{2} - w_{i} =
\|\mathbf{x}_{j}-\mathbf{o}_{f}\|^{2} - w_{j} =
\|\mathbf{x}_{k}-\mathbf{o}_{f}\|^{2} - w_{k}\,.
\end{equation}
Following suitable algebraic manipulation\footnote{{\reviewerA While expressions for cell orthocentres can be written in various forms, note that the linear system shown here does not involve absolute coordinate/weight values, but is instead written in terms of relative \textit{differences}. As per Shewchuk \cite{shewchuk1999lecture}, such a formulation is designed to improve numerical robustness when computations are executed in finite-precision arithmetic.}}, these expressions lead to the following system of linear equations for the point $\mathbf{o}_{f}$
\begin{gather}
\begin{bmatrix}
\delta x_{ij} & \delta y_{ij} \\[.75ex]
\delta x_{ik} & \delta y_{ik}
\end{bmatrix}
\begin{bmatrix}
\Delta x \\[.75ex]
\Delta y
\end{bmatrix} =
\frac{1}{2} 
\begin{bmatrix}
\delta x_{ij}^{2} + \delta y_{ij}^{2} - \delta w_{ij} \\[.75ex]
\delta x_{ik}^{2} + \delta y_{ik}^{2} - \delta w_{ik}
\end{bmatrix}\,,
\\[1.33ex]
\mathbf{o}_{f} = \mathbf{x}_{i} + (\Delta x, \Delta y)\,,
\end{gather}
where $\delta(\cdot)_{ij}$ denotes the difference $(\cdot)_{j} - (\cdot)_{i}$. Note that in unweighted configurations with $w_{i} = w_{j} = w_{k} = 0$, the point $\mathbf{o}_{f}$ is equivalent to the centre of the circumscribing ball $\mathbf{c}_{f}$ associated with $\tau_{i}$. Noting again that the expression for $\mathbf{o}_{f}$ is a function of the local weight differences $\delta w_{ij}$, $\delta w_{ik}$, equally-weighted triangles with $w_{i} = w_{j} = \beta \in \mathbb{R}$ reduce to $\mathbf{o}_{f} = \mathbf{c}_{f}$. Such considerations demonstrate that Regular-Power structures differ from conventional Delaunay-Voronoi pairs only in cases where there is a non-uniform distribution of weights throughout the grid.

{
\reviewerA

\section{Discretisation errors in a model co-volume scheme}
\label{section_numerical_error}

To motivate discussions of optimality in various primal-dual grid configurations, an analysis of the numerical errors associated with a `model' co-volume scheme is presented, based on a mimetic-type discretisation of a nonlinear momentum equation and a hyperbolic transport law
\begin{gather}
\partial_{t}\mathbf{u} + (\mathbf{u}\cdot\nabla)\,\mathbf{u} = -\nabla \phi\,,
\label{eqn_momentum_0}
\\[2ex]
\partial_{t} \psi + \nabla \cdot (\mathbf{u}\, \psi) = 0\,.
\label{eqn_continuity_0}
\end{gather}

Here $\mathbf{u}$ is a two-dimensional velocity field $\mathbf{u} = (u,v)$, $\psi$ is a passive scalar distribution advected with the flow and $\phi$ is an unspecified potential function. $\partial_{t}$ denotes a derivative with respect to time $t$, and $\nabla = (\partial_{x},\partial_{y})$. Following, for example, Ringler et al \cite{ringler2010unified}, the momentum equation (\ref{eqn_momentum_0}) can be expressed in so-called  \textit{vector-invariant} form via suitable manipulation of the following vector identity
\begin{gather}
(\mathbf{u}\cdot\nabla)\, \mathbf{u} \rightarrow (\nabla \times \mathbf{u}) \times \mathbf{u} + \tfrac{1}{2}\,\nabla (\mathbf{u} \cdot \mathbf{u})\,.
\label{eqn_vector_identity}
\end{gather}

Substituting (\ref{eqn_vector_identity}) into (\ref{eqn_momentum_0}), and defining the absolute vorticity $\xi = \hat{\mathbf{k}}\cdot\nabla\times\mathbf{u}$ and the kinetic energy $K = \tfrac{1}{2}\,(\mathbf{u}\cdot\mathbf{u})$, the vector-invariant form of the model system can be written as
\begin{gather}
\partial_{t}\mathbf{u} + \xi \mathbf{u}^{\perp} = -\nabla \phi - \nabla K\,,
\label{eqn_momentum_1}
\\[2ex]
\partial_{t} \psi + \nabla \cdot (\mathbf{u}\, \psi) = 0\,,
\label{eqn_continuity_1}
\end{gather}

where $\mathbf{u}^{\perp} = (-v,u)$ is the perpendicular velocity field. As per, for example, Ringler et al \cite{ringler2010unified,ringler2013multi}, such a formulation allows for the construction of mimetic discretisations that explicitly conserve rotational and energetic invariants associated with the system, and is typically preferred for the discretisation of geophysical flows as a result.

\begin{figure*}[t]
  \centering
  \includegraphics[width=.9125\textwidth]{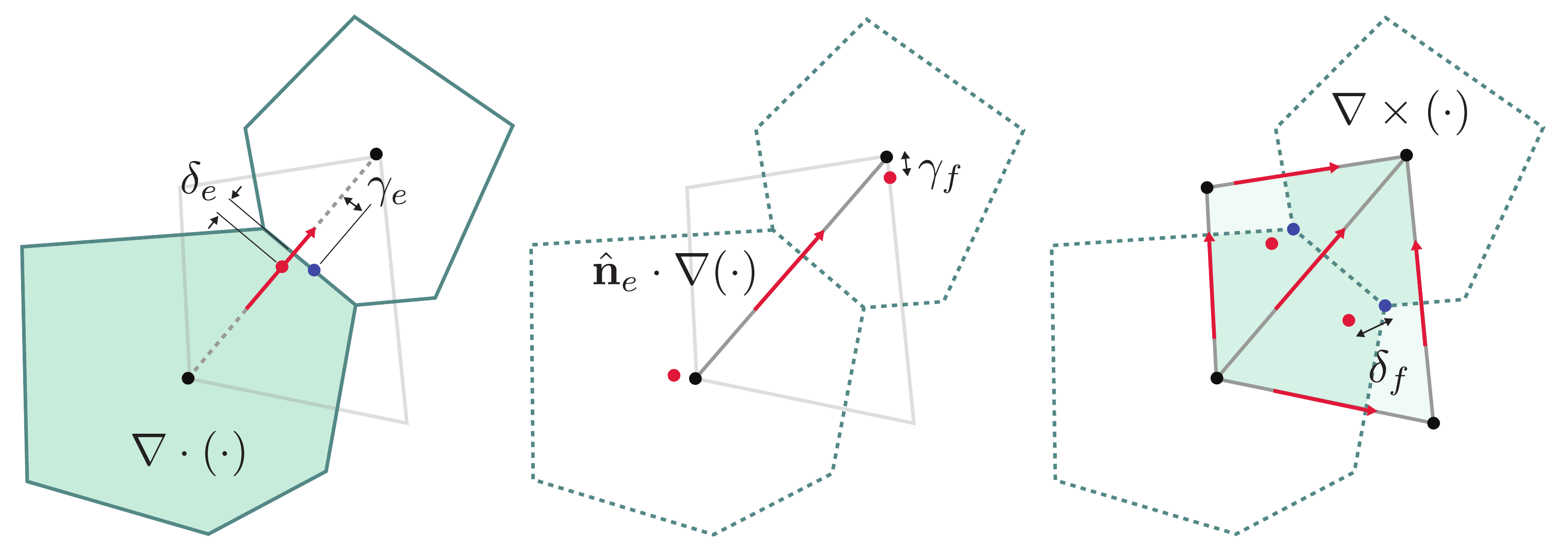}
  
  \caption{Mimetic operators defined on a staggered unstructured grid, showing (a) the reconstruction of edge fluxes for the discrete divergence operator defined on the dual cells, (b) evaluation of a discrete normal gradient operator associated with primal edges, and (c) computation of discrete vorticity on the primal cells. The various edge- and face-centred primal-dual `defect' terms are illustrated, consistent with the discretisation errors associated with each operator.
  }
  \label{figure_grid_error}
\end{figure*}

\subsection{A staggered mimetic formulation}

\medskip

Following Ringler et al \cite{ringler2010unified}, the coupled system (\ref{eqn_momentum_1})--(\ref{eqn_continuity_1}) can be discretised via a mimetic strategy, in which a set of normal velocity components $(\mathbf{u}\cdot\hat{\mathbf{n}})_{j}$ are positioned on and aligned with the edges of the primal triangulation, conserved cell-mean quantities $\bar{\psi}_{i} = {|d_{i}|}^{-1} \int_{d_{i}} \psi\,\, \mathrm{d}A$ are located within the cells $d_{i}$ of the dual grid and a discrete vorticity distribution $\bar{\xi}_{k} = {|\tau_{k}|}^{-1} \int_{\tau_{k}} \nabla \times \mathbf{u}\,\, \mathrm{d}A$ is computed on the triangles $\tau_{k}$ in the primal complex. As per the discussions presented in Section~\ref{section_grid_constraints}, such a formulation can be interpreted as an unstructured variant of the well-known Arakawa-type C-grid scheme, with the nonlinear momentum equation (\ref{eqn_momentum_1}) expressed in terms of mimetic finite-difference operators, and the continuity equation (\ref{eqn_continuity_1}) discretised following a conservative finite-volume approach. See Figure~\ref{figure_primal_dual_TRSK} for additional details. 

Here, the numerical errors associated with such a formulation are analysed, and such analysis is used to inform the design of optimised primal-dual grid configurations. The effect of various geometrical offsets, here termed `defects', are considered: 
\begin{itemize}\itemsep -1pt
\item $\delta_{f}$ | the offset between dual vertices and primal cell centres.
\item $\delta_{e}$ | the offset between dual edges and primal edge midpoints. 
\item $\gamma_{f}$ | the offset between primal vertices and dual cell centres. 
\item $\gamma_{e}$ | the offset between primal edges and dual edge midpoints. 
\end{itemize}
`Perfectly' regular primal-dual pairs consisting of equilateral triangles and uniform hexagonal dual cells satisfy $\delta_{f} = \delta_{e} = 0$ and $\gamma_{f} = \gamma_{e} = 0$. In subsequent sections, the impact of `imperfect' grids where $\delta_{f}>0$, $\delta_{e}>0$, $\gamma_{f}>0$ and $\gamma_{e}>0$ are analysed in detail.

\subsection{Errors associated with $\nabla \cdot (\cdot)$}

\medskip

A finite-volume approximation for the discrete divergence operator in the continuity equation (\ref{eqn_continuity_1}) can be expressed as a summation of cell-edge fluxes. Integrating over a given cell $d_{i}$ in the dual grid
\begin{gather}
\label{eqn_discrete_div}
\int_{d_{i}} \nabla \cdot (\mathbf{u}\, \psi)\,\, \mathrm{d}A = 
\oint_{\partial d_{i}} (\mathbf{u}\cdot\hat{\mathbf{n}})\, \psi\,\, \mathrm{d}s \simeq
\sum_{e=1}^{n} \int_{e}\, (\mathbf{u}\cdot\hat{\mathbf{n}})_{e}\, \psi_{e}\,\, \mathrm{d}l\,,
\end{gather} 

where the edge index $e$ iterates over the edges of the cell $d_{i}$. Use of a suitable quadrature rule to evaluate the integrals over edges in (\ref{eqn_discrete_div}) leads to a fully discrete method. In practice, (\ref{eqn_discrete_div}) is typically discretised using a midpoint rule, with $(\mathbf{u}\cdot\hat{\mathbf{n}})_{e}$ and $\psi_{e}$ evaluated at a single quadrature point on each edge $e$. Given a perfectly regular tessellation, with $\delta_{e} = \gamma_{e} = 0$, such a formulation achieves second-order accuracy, with the normal velocity components coincident with the quadrature points positioned at the midpoints of the dual edges. Numerical accuracy is degraded as grid distortion increases, with $\delta_{e} > 0$, $\gamma_{e} > 0$ leading to misalignment between $(\mathbf{u}\cdot\hat{\mathbf{n}})_{e}$ and the associated quadrature points. See Figure~\ref{figure_grid_error}a for details. Note also that the discretisation becomes \textit{non-interpolatory} when the grid is not well-centred, with adjacent primal and dual edges failing to intersect in such cases.

\subsection{Errors associated with $\nabla(\cdot)$}

\medskip

A finite-difference approximation for the discrete gradient operator in the momentum equation (\ref{eqn_momentum_1}) can be formed by exploiting the orthogonality of the underlying primal-dual tessellation. Expressing the momentum equation in terms of normal velocity components
\begin{gather}
\partial_{t}(\mathbf{u}\cdot\hat{\mathbf{n}}_{e}) + \hat{\mathbf{n}}_{e} \cdot (\xi \mathbf{u}^{\perp}) = -\hat{\mathbf{n}}_{e} \cdot \nabla \phi - \hat{\mathbf{n}}_{e} \cdot \nabla K\,,
\label{eqn_momentum_2}
\end{gather}
 
it is clear that an expression for the normal gradients, rather than full gradient vector, is required to be evaluated on each primal edge $e$. Noting that the normal velocity components are positioned at the midpoints of primal edges, the two-point finite-difference
\begin{gather}
\label{eqn_discrete_grad}
\hat{\mathbf{n}}_{e} \cdot \nabla \phi \simeq {l_{e}}^{-1}\,(\phi_{2} - \phi_{1})\,,
\end{gather}

where $l_{e}$ is the length of the primal edge $e$, and $\phi_{1}$, $\phi_{2}$ are values of $\phi$ centred at the two adjacent primal vertices, achieves second-order accuracy. See Figure~\ref{figure_grid_error}b for detail. Additional consideration is needed when $\phi_{1}$, $\phi_{2}$ are not specified directly at primal vertices, but are instead reconstructed from cell-mean data stored on the dual cells. Such an arrangement is consistent with, for example, evaluation of the pressure gradient force in geophysical flows. In such cases, second-order accuracy is maintained only when the tessellation is centroidal, with primal vertices coincident with dual cell centres, such that $\gamma_{f} = 0$.

\subsection{Errors associated with $\nabla \times (\cdot)$, $\mathbf{u}^{\perp}$, $\tfrac{1}{2}(\mathbf{u}\cdot\mathbf{u})$}

\medskip

A discrete approximation for the nonlinear terms in the momentum equation (\ref{eqn_momentum_1}) requires a sequence of composite numerical operations; reconstructing the full velocity vector field from its normal components and evaluating a discrete vorticity distribution. Various vector field reconstruction techniques exist, including Perot's method \cite{PEROT200058}, least-squares interpolation and mixed finite-element formulations \cite{PEIXOTO2014185}, or the TRSK operators introduced in Thurburn et al \cite{thurburn2009}. As per the analysis of Peixoto \cite{peixoto2016accuracy}, these methods typically require perfectly regular tessellations (i.e.~$\delta_{e} = \gamma_{e} = 0$) to achieve higher-order accuracy. The accuracy of the reconstructed perpendicular velocity field $\mathbf{u}^{\perp}$ and the kinetic energy distribution $K = \tfrac{1}{2}(\mathbf{u}\cdot\mathbf{u})$ are constrained accordingly.

A finite-volume type vorticity distribution can be reconstructed over the cells in the primal mesh by exploiting the alignment of the discrete velocity components with primal edges
\begin{gather}
\label{eqn_discrete_curl}
|\tau_{k}|\, \bar{\xi}_{k} = \int_{\tau_{k}} \nabla \times \mathbf{u}\,\, \mathrm{d}A = 
\oint_{\partial \tau_{k}} (\mathbf{u}\cdot \hat{\mathbf{t}})\,\, \mathrm{d}s \simeq
\sum_{e=1}^{3} \int_{e}\, (\mathbf{u}\cdot \hat{\mathbf{t}})_{e}\,\, \mathrm{d}l\,.
\end{gather}

Here, $\bar{\xi}_{k}$ is the integral-mean vorticity induced within a given primal cell $\tau_{k}$ and $\hat{\mathbf{t}}_{e}$ is the unit tangent vector associated with a primal edge $e$, where, due to the orthogonality of the grid, $\hat{\mathbf{t}}_{e} = \hat{\mathbf{k}} \times \hat{\mathbf{n}}_{e}$. Noting that the tangential velocity components $(\mathbf{u}\cdot \hat{\mathbf{t}})_{e}$ are positioned at the midpoints of the primal edges, the cell-mean vorticity $\bar{\xi}_{k}$ can be evaluated to second-order accuracy on the primal cells $\tau_{k}$ by discretising (\ref{eqn_discrete_curl}) using a midpoint quadrature rule. Evaluation of the rotational contribution to the momentum equation $\hat{\mathbf{n}}_{e} \cdot (\eta\,\mathbf{u}^{\perp})$ is completed by construction of a suitable interpolation operator, expressing the vorticity at the primal edge midpoints in terms of the adjacent primal cell-mean values. Following Ringler et al \cite{ringler2010unified}, a simple arithmetic average $\xi_{e} = \tfrac{1}{2}(\bar{\xi}_{1}+\bar{\xi}_{2})$, with $\bar{\xi}_{1}$, $\bar{\xi}_{2}$ the cell-mean values adjacent to a given primal edge $e$, achieves second-order accuracy when the primal-dual pair is perfectly centred. Such configurations require $\delta_{f} = 0$. See Figure~\ref{figure_grid_error}c for detail.

\subsection{Implications for primal-dual grid staggering}

\medskip

Clearly, the magnitude of the discretisation error associated with various unstructured mimetic numerical schemes \cite{ringler2010unified,ringler2013multi,thurburn2009,KORN2017525,KORN2017156} is a strong function of the regularity of the underlying primal-dual grid, with higher-order accuracy achieved only for uniform tessellations with $\delta_{f} = \delta_{e} = 0$ and $\gamma_{f} = \gamma_{e} = 0$. In general, imperfections in the grid staggering induce geometrical offsets between the edges and faces of adjacent primal and dual cells, leading to a degradation in the accuracy of the various discrete divergence, gradient and/or rotational terms defined by the model system (\ref{eqn_momentum_1})--(\ref{eqn_continuity_1}), as detailed previously. The minimisation of such grid defects is highly desirable as a result.

By construction, conventional Delaunay-Voronoi pairs satisfy $\delta_{e} = 0$, guaranteeing that Voronoi edges cross primal edges at their midpoints. Additionally, so-called Centroidal Voronoi Tessellations (CVTs) ensure that primal vertices are positioned at the centre of their adjacent dual Voronoi cells, corresponding to $\gamma_{f} \rightarrow 0$. Such structures, though, do not explicitly bound the offsets between primal edges and Voronoi edge midpoints, $\gamma_{e}$, or the offsets between dual vertices and primal element centroids, $\delta_{f}$. Despite consisting of very high quality primal triangulations, high-quality centroidal Delaunay-Voronoi grids do not necessarily contain optimal dual structure, and are often not well-centred as a result. In the following sections, a generalised primal-dual formulation based on the Regular-Power pair is pursued, in which the distribution of vertex weights is carefully selected to minimise the various primal-dual defect terms. Specifically, by relaxing the $\delta_{e} = 0$ constraint associated with conventional Voronoi structures, higher-quality Power diagrams are sought; designed to more evenly minimise the full set of defects $\delta_{f}$, $\delta_{e}$, $\gamma_{f}$ and $\gamma_{e}$. These optimised primal-dual tessellations are expected to lead to improved performance of various staggered unstructured numerical formulations as a result.

}

\section{Weight selection}
\label{section_dual_optim}

Before addressing the generation of optimal primal-dual tessellations in full, the task of static \textit{weight selection} is explored; seeking to choose a set of scalar weights $W \in \mathbb{R}$ that optimise the geometry of the dual grid $\mathcal{D}(\mathbf{X},W)$ associated with a given set of points $\mathbf{X} \in \mathbb{R}^{d}$. Here, the two-dimensional problem is considered, where $\mathbf{X} \in \mathbb{R}^{2}$ and the primal-dual tessellation $(\mathcal{T},\mathcal{D})$ is a pair of two-dimensional complexes, consisting of collections of triangles and polygons, respectively. In the present work, the weight selection task is cast as an optimisation problem; seeking to find a set of weights that maximise a given dual grid \textit{quality} metric.

\subsection{A dual grid quality metric}

\begin{figure*}[t]
  \centering
  \includegraphics[width=.7375\textwidth]{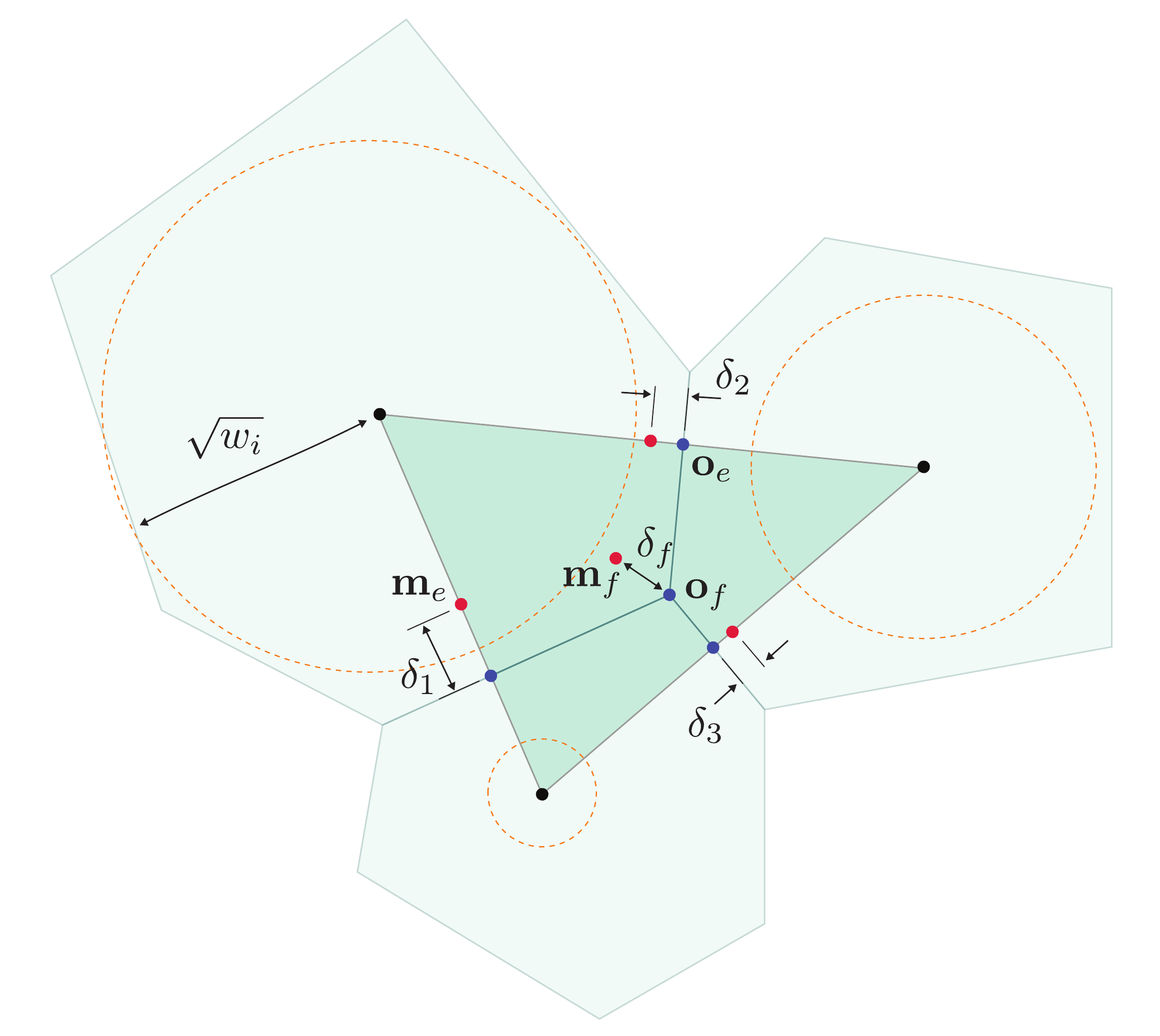}
  
  \caption{Anatomy of the dual cost metric (\ref{eqn_dual_quality}), showing the contributions to $\mathcal{Q}_{i}^{\mathcal{D}}(\mathbf{X},W)$ for a single triangle $\tau_{i} \in \mathcal{T}(\mathbf{X},W)$. Here, the triangle and edge centroids $\mathbf{m}_{f}$ and $\mathbf{m}_{e}$ are drawn in red, and the associated weighted orthocentres $\mathbf{o}_{f}$ and $\mathbf{o}_{e}$ are drawn in blue. Non-uniform vertex weights $w_{i}$ are applied at the vertices of $\tau_{i}$. The dual quality metric is a function of the local face- and edge-centred \textit{defects} $\delta_{f} = \|\mathbf{m}_{f}-\mathbf{o}_{f}\|$ and $\delta_{e} = \|\mathbf{m}_{e}-\mathbf{o}_{e}\|$; representing the geometrical offset between adjacent faces in the generalised pair $(\mathcal{T},\mathcal{D})$.
  }
  \label{figure_dual_cost}
\end{figure*}

\medskip

Given a triangle $\tau_{i}$ in the weighted primal mesh $\mathcal{T}(\mathbf{X},W)$, a local dual grid quality metric $\mathcal{Q}^\mathcal{D}_{\,i}(\mathbf{X},W)$ is proposed; designed to measure the relative geometrical \textit{defect} between polygonal cells in the orthogonal dual $\mathcal{D}(\mathbf{X},W)$ and an ideal centroidal configuration
{\reviewerA
\begin{gather}
\label{eqn_dual_quality}
\mathcal{Q}^\mathcal{D}_{\,i}(\mathbf{X},W) = 
\beta_{f}\underbrace{\left(1 - \left(\frac{\delta_{f}}{\bar{l}_{f}}\right)^{2}\right)}_{\text{`defect' at face}} + \, 
\beta_{e}\underbrace{\left(\frac{1}{3}\,\sum_{e=1}^{3}\, 1 - \left(\frac{\delta_{e}}{l_{e}}\right)^{2}\right)}_{\text{mean `defect' at edges}}\,,
\\[2ex]
\text{with}\quad
\delta_{f} = \|\mathbf{o}_{f}-\mathbf{m}_{f}\|\,,
\quad
\delta_{1,2,3} = \|\mathbf{o}_{1,2,3}-\mathbf{m}_{1,2,3}\|\,.
\end{gather}}

Here, $\mathbf{o}_{f}$, $\mathbf{o}_{e}$ are the weighted orthocentres associated with the face and edge segments of the triangle $\tau_{i}$, the points $\mathbf{m}_{f}$, $\mathbf{m}_{e}$ are the corresponding face and edge centroids, and $\bar{l}_{f} = \frac{1}{3}\sum l_{e}$ is a local characteristic length, taken as the mean of the incident edge lengths $l_{e}$. Such a scaling serves to non-dimensionalise the various defect terms in (\ref{eqn_dual_quality}), and results in a metric $\mathcal{Q}^\mathcal{D}_{\,i}(\mathbf{X},W)$ that is \textit{scale-invariant}. {\reviewerA The linear coefficients $\beta_{f}$ and $\beta_{e}$ weight the face- and edge-centred contributions to $\mathcal{Q}^\mathcal{D}_{\,i}(\mathbf{X},W)$. In this study, a simple average is employed, with $\beta_{f} = \beta_{e} = \frac{1}{2}$, indicating that equal `importance' is assigned to both face- and edge-centred defects.} See Figure~\ref{figure_dual_cost} for additional detail. 

The metric $\mathcal{Q}^\mathcal{D}_{\,i}(\mathbf{X},W)$ is a smooth function of the geometry of the triangle $\tau_{i}$ and the weights centred at its vertices. A maximum value $\mathcal{Q}^\mathcal{D}_{\,i}(\mathbf{X},W) = 1$ is attained when the dual cell is optimally staggered with respect to the triangle $\tau_{i}$; spanning between its centroid and edge midpoints, such that $\mathbf{o}_{f} = \mathbf{m}_{f}$ and {\reviewerB$\mathbf{o}_{e} = \mathbf{m}_{e}$}. Conversely, $\mathcal{Q}^\mathcal{D}_{\,i}(\mathbf{X},W) \rightarrow 0$ as the relative face- and/or edge-centred defects increase as the local primal-dual staggering becomes less centroidal. The function $\mathcal{Q}^\mathcal{D}_{\,i}(\mathbf{X},W)$ aims to provide a combined measure of the quality of the staggering between primal and dual grid cells, with the first term in (\ref{eqn_dual_quality}) accounting for the defect between the dual grid vertices and triangle centroids, $\delta_{f}$, and the second term the mean defect between dual grid edges and triangle edge midpoints, $\delta_{e}$. Such considerations are designed to minimise the numerical error of the model co-volume discretisation scheme described previously, where a mutually centroidal staggering between the primal and dual grid cells, vertices and edges is optimal. 

{\reviewerA Optimisation of $\mathcal{Q}^\mathcal{D}_{\,i}(\mathbf{X},W)$ is expected to lead to configurations with $\delta_{f} \rightarrow 0$ and $\delta_{e} \rightarrow 0$. When combined with a CVT-like optimisation to also improve the quality of the primal tessellation (i.e.~$\gamma_{f} \rightarrow 0$), the combined scheme is expected to generate very high-quality staggered unstructured grids, exceeding the capabilities of conventional Delaunay-Voronoi pairs. In particular, direct minimisation of $\delta_{f}$ is expected to lead to the construction of higher quality dual structures and the generation of well-centred tessellations.

In the context of geophysical fluid dynamics, a detailed error analysis for the mimetic formulation of Ringler et al \cite{ringler2010unified,ringler2013multi} is presented by Peixoto in \cite{peixoto2016accuracy}, and, in conjunction with the analysis presented here in Section~\ref{section_numerical_error}, shows that the various defects between primal element centroids and dual vertices, and primal and dual edge midpoints are significant sources of overall numerical error; limiting the order of accuracy of the underlying numerical operators. The dual quality metric $\mathcal{Q}_{i}^{D}(\mathbf{X},W)$ is intended as a `penalty' operator for such primal-dual offsets, and is expected to improve the accuracy of such staggered unstructured discretisation schemes.
}


\subsection{Selection of locally-optimal weights}

\medskip

Given a weighted primal triangulation $\mathcal{T}(\mathbf{X},W)$, the task of constructing a high-quality orthogonal dual $\mathcal{D}(\mathbf{X},W)$ can be cast as an optimisation problem  
\begin{equation}
\label{eqn_dual_optim_problem}
\text{find } {\reviewerB W \subset \mathbb{R}}, \text{ such that } {\reviewerB \min \mathcal{Q}^\mathcal{D}_{\,i}(\mathbf{X},W)}\, \,\forall\,\, \tau_{i}\in\mathcal{T}(\mathbf{X},W) \text{ is maximised}.
\end{equation}
In general, (\ref{eqn_dual_optim_problem}) is a global, non-convex optimisation problem, with the weight at each vertex $w_{i}\in\mathcal{T}(\mathbf{X},W)$ contributing to the non-linear metric $\mathcal{Q}^\mathcal{D}_{\,k}(\mathbf{X},W)$, computed for all adjacent triangles $\tau_{k}\in\mathcal{T}(\mathbf{X},W)$. Rather than seeking solutions to this global problem directly, a simpler, locally-optimal approach is pursued here, inspired by the constrained, gradient-ascent type methods introduced for primal mesh smoothing in, for example, work by Freitag and Ollivier-Gooch \cite{freitag1997tetrahedral} and Klinger and Shewchuk \cite{klingner2008aggressive}. 

Specifically, the solution of a sequence of local, decoupled problems is considered, with a local optimisation of each given vertex weight $w_{i}$ computed incrementally. Noting that each weight $w_{i}$ has local influence, contributing to the metrics $\mathcal{Q}^\mathcal{D}_{\,k}(\mathbf{X},W)$ for the triangles $\tau_{k}\in\mathcal{T}(\mathbf{X},W)$ adjacent to {\reviewerB $\mathbf{x}_{i}$} only, a steepest-ascent type update is employed, with
\begin{equation}
\label{eqn_dual_optim_update}
w_{i}^{n+1} = w_{i}^{n} + \Delta_{i}^{m}\, {\reviewerB v_{i}^{n}}\,,
\end{equation}
where
\begin{equation}
{\reviewerB v_{i}^{n}} = \frac{d}{dw_{i}}\, \mathcal{Q}^\mathcal{D}_{\,j}(\mathbf{X},W)
\quad \text{and}\, \quad
j = \operatorname{argmin}_{k} \mathcal{Q}^\mathcal{D}_{\,k}(\mathbf{X},W) \,\,\forall \text{ adj. } \tau_{k}\in\mathcal{T}(\mathbf{X},W)\,.
\end{equation}
Here, the index $k$ is taken as a loop over the primal triangles $\tau_{k} \in\mathcal{T}(\mathbf{x},w)$ incident to {\reviewerB$\mathbf{x}_{i}$}. The scalar step length $\Delta_{i}^{m} \in \mathbb{R}^{+}$ is computed via a line search along the gradient ascent vector {\reviewerB$v_{i}$} and, in this study, is taken as the first value that leads to an improvement in the worst-case incident-quality metric $\mathcal{Q}^\mathcal{D}_{\,j}(\mathbf{X},W)$. Note that the ascent vector {\reviewerB$v_{i}^{n}$} is {\reviewerB taken} as the gradient of the \textit{worst} incident quality metric, rather than some mean measure, and that the scheme is biased toward providing updates in a \textit{worst-first} manner as a result.

A local line search is performed via a simple bisection strategy, iteratively reducing the step-length such that $\Delta_{i}^{m} = (\frac{1}{2})^{m}\, \Delta\bar{w}$, where $m$ is the local line search iterate. The initial guess $\Delta\bar{w}$ is computed by considering a first-order Taylor expansion in local grid quality metrics
\begin{equation}
\label{eqn_dual_optim_step}
\mathcal{Q}^\mathcal{D}_{\,j}(\mathbf{X},W) + \Delta\bar{w}\, \frac{d}{dw_{i}} \, \mathcal{Q}^\mathcal{D}_{\,j}(\mathbf{X},W) \leq \bar{\mathcal{Q}}^\mathcal{D}_{\,k}(\mathbf{X},W)\,.
\end{equation} 
Here, $\bar{\mathcal{Q}}^\mathcal{D}_{\,k}(\mathbf{X},W)$ is a taken as a mean quality value over the local incident set $\tau_{k} \in \mathcal{T}(\mathbf{X},W)$, with the corresponding $\Delta\bar{w}$ an estimate of the weight perturbation required to improve the worst metric until it is equal to the local mean measure. A limited line search is employed, testing iterations $m \leq 5$ until a successful step is found. If no such improvement is identified, the weight is left unchanged, with $w_{i}^{n+1} = w_{i}^{n}$. Note that such a procedure is guaranteed to result in \textit{monotone} improvements in the grid quality function $\mathcal{Q}^{\mathcal{D}}(\mathbf{X},W)$; ensuring that the worst-case grid quality metric in the incident set is never decreased. This behaviour is here referred to as a \textit{quality-preserving} update strategy.

The local scheme described in (\ref{eqn_dual_optim_update})--(\ref{eqn_dual_optim_step}) can be used to construct pseudo-optimal solutions to the global problem (\ref{eqn_dual_optim_problem}) via iteration. In this work, a weakly stochastic, Gauss-Seidel type update is employed, consisting of a series of global sweeps over the full set of vertex weights, with each weight updated incrementally using the local scheme described in (\ref{eqn_dual_optim_update})--(\ref{eqn_dual_optim_step}). Within each outer iteration, vertices are visited in a randomised order; seeking to reduce the likelihood of the iteration becoming trapped in order-induced local optima. Consistent with a Gauss-Seidel type philosophy, updated weight values are used in subsequent metric evaluations as soon as they are available, such that $w_{i-1}^{n+1}$ contributes to the calculation of $\mathcal{Q}^\mathcal{D}_{\,i}(\mathbf{X},W)$ for all $i > i-1$.

While such a procedure is not guaranteed to find a globally optimal solution to (\ref{eqn_dual_optim_problem}), it does inherit the \textit{quality-preserving} nature of the local gradient-ascent scheme; ensuring that the worst-case grid quality metrics are improved monotonically. Noting that the accuracy and performance of a numerical discretisation scheme typically scales with the worst quality elements in a grid, rather than some mean measure, it is suggested that such behaviour is highly desirable; consistent with the development of so-called \textit{smart} node smoothing schemes for primal meshes, as per \cite{freitag1997tetrahedral,klingner2008aggressive}.

\section{Optimisation of primal-dual grids}
\label{section_primal_dual_optim}

Building on the weight-selection procedure introduced in Section~\ref{section_dual_optim}, the full primal-dual grid generation algorithm is presented; seeking to find a set of vertex positions $\mathbf{X}$ and associated vertex weights $W$ that induce a primal-dual pair with optimal geometrical and topological characteristics. Consistent with the weight-selection strategy presented previously, the grid generation task is cast as an optimisation problem; seeking to maximise a pair of primal-dual grid quality metrics. All primal and dual optimisation operations are designed to be \textit{quality-preserving}; ensuring that improvements in the local, worst-case incident grid quality metrics $\mathcal{Q}^{\mathcal{T}}(\mathbf{X})$ and/or $\mathcal{Q}^{\mathcal{D}}(\mathbf{X},W)$ are achieved where possible. Such behaviour is designed to ensure the overall robustness of the scheme; resulting in a coupled optimisation strategy designed to improve the worst-case elements in a given primal-dual grid.

\subsection{A primal grid quality metric}

\medskip

Given a triangle $\tau_{i}$ in the weighted primal mesh $\mathcal{T}(\mathbf{X},W)$, the \textit{area-length} ratio \cite{Parthasarathy94} is employed as a local primal grid quality metric $\mathcal{Q}_{i}^{\mathcal{T}}(\mathbf{X})$; designed to measure the relative distortion and irregularity of the primal grid cells. 
\begin{equation}
\label{eqn_tria_quality}
\mathcal{Q}_{i}^{\mathcal{T}}(\mathbf{X}) = \frac{4 \sqrt{3}}{3} \frac{A_{i}}{\|\mathbf{e}\|_{\mathrm{rms}}^{2}}\,.
\end{equation}
Here, $A_{i}$ is the signed area of the triangle $\tau_{i}$ and $\|\mathbf{e}\|_{\mathrm{rms}}$ is the associated root-mean-square edge length. The area-length ratio is a robust, scalar measure of triangle shape quality, and is normalised to achieve a score of $+1$ for ideal elements. The area-length ratio decreases with increasing cell distortion, achieving a score of $+0$ for degenerate elements and $-1$ for `tangled' elements with reversed orientation.

{\reviewerA As per Shewchuk \cite{shewchuk2002good}, the area-length metric is a measure of the condition number of various numerical operators and transformation matrices expressed on the primal grid. Compared to other primal mesh quality criteria, it is numerically well-behaved, is relatively cheap to evaluate and will detect all classes of low quality simplexes. The area-length metric and its generalisations have been  successfully employed in a number of studies, including, for example, work by Klinger and Shewchuck \cite{klingner2008aggressive}, Gosselin and Olliver-Gooch \cite{Gosselin2011}, as well as previous investigations by the author \cite{engwirda2017jigsaw}.}

\subsection{Selection of locally-optimal vertex positions}

\medskip

Considering, firstly, the \textit{geometrical} optimality of the primal grid $\mathcal{T}(\mathbf{X},W)$, a \textit{mesh-smoothing} procedure is undertaken; seeking to reposition vertices to improve both the primal grid quality metrics $\mathcal{Q}^{\mathcal{T}}(\mathbf{X})$ and conformance to a user-defined mesh-spacing function $\bar{h}(\mathbf{x}): \mathbb{R}^{2} \rightarrow \mathbb{R}^{+}$, defining the distribution of desired primal edge lengths throughout the domain. A hybrid, two-pass strategy is pursued; employing a variation of the Optimal Delaunay Triangulation (ODT) scheme of Chen et al \cite{chen2004optimal,chen2011efficient} as a first-pass update, and falling back to a constrained, gradient-ascent type approach when required.

\subsubsection{`Optimal' weighted triangulations}

\medskip

Given a vertex $\mathbf{x}_{i}$ in the primal mesh $\mathcal{T}(\mathbf{X},W)$, an initial update is attempted using a variation on the ODT strategy of Chen et al \cite{chen2004optimal,chen2011efficient}. In the ODT formulation, primal vertices are repositioned such that an element-wise \textit{energy-functional} is minimised; leading to the optimal piecewise linear reconstruction of quadratic functionals over an associated Delaunay triangulation $\mathcal{T}(\mathbf{X}) = \operatorname{Del}(\mathbf{X})$. The ODT scheme is known to be an effective Delaunay-based mesh optimisation strategy; leading to high quality Delaunay triangulations and Voronoi diagrams that can be adapted to general, spatially-dependent mesh-density functions $\bar{\rho}(\mathbf{x}): \mathbb{R}^{2} \rightarrow \mathbb{R}^{+}$ \cite{chen2004optimal,chen2011efficient}.

In the conventional ODT formulation, primal vertices are repositioned to the weighted mean of adjacent triangle circumcentre coordinates; exploiting the orthogonal structure of the standard Delaunay-Voronoi tessellation. In the current work, such a strategy is modified to incorporate the generalised, weighted primal-dual structure pursued throughout; replacing the triangle circumcentres with the associated element orthocentres. This modification leads to a scheme in which primal vertex positions are updated as a weighted sum of the adjacent dual vertex coordinates; maintaining conceptual consistency with the conventional, unweighted ODT strategy. The resulting scheme is denoted Optimal Weighted Triangulation (OWT)\footnote{Though additional methods are not pursued in the present study, note that a generalisation of a range of existing variational mesh optimisation schemes, including the popular Centroidal Voronoi Tessellation (CVT) \cite{du1999centroidal,du2002grid} could be achieved through similar modifications to the definition of dual vertices; replacing element circumcentre coordinates with those of the associated weighted orthocentre.} throughout. {\reviewerB Adapting the ODT-style update strategy of Chen and Holst \cite{chen2011efficient}, primal vertices are repositioned such that}
\begin{equation}
\label{eqn_tria_optim_update_1}
\mathbf{x}_{i}^{n+1} = \left(1 - \Delta_{i}^{m}\right)\,\mathbf{x}_{i}^{n} + \Delta_{i}^{m} \sum_{\tau_{j} \in\, {*}_{i}} \frac{|\tau_{j}|_{\bar{h}}}{|{*}_{i}|_{\bar{h}}}\,\, \mathbf{o}_{j}\,.
\end{equation}
Here, ${*}_{i}$ denotes the \textit{star} of $\mathbf{x}_{i}$ | the local set of elements $\tau_{j} \in \mathcal{T}(\mathbf{X},W)$ adjacent to $\mathbf{x}_{i}$. $|\tau_{j}|_{\bar{h}}$ denotes the \textit{spacing-weighted} area of the triangle $\tau_{j}$, and $|{*}_{i}|_{\bar{h}}$ the summation of such terms over the set ${*}_{i}$. The points $\mathbf{o}_{j}$ are the orthocentres associated with the triangles $\tau_{j}$ and $\Delta_{i}^{m}$ is a relaxation factor, computed via a local line search. In this study, the weighted-area factors are computed using a 1-point quadrature rule
\begin{equation}
|\tau_{j}|_{\bar{h}} = 
\int_{\tau_{j}} \frac{1}{\bar{h}(\mathbf{x})^{2}}\,\, \mathrm{d}A \simeq \frac{A_{j}}{\bar{h}_{j}^{2}}
\quad\text{with}\quad
\bar{h}_{j} = \frac{1}{3}\sum_{x_{k} \,\in\, \tau_{j}} \bar{h}(\mathbf{x}_{k})\,.
\end{equation}
Conventionally (see, for example \cite{chen2011efficient}), such terms are computed with respect to an imposed \textit{mesh-density} function $\bar{\rho}(\mathbf{x})$ rather than a \textit{mesh-spacing} function $\bar{h}(\mathbf{x})$. Here, the relation $\bar{\rho}(\mathbf{x}) = \nicefrac{1}{\bar{h}(\mathbf{x})^{d}}$ is used to express density in terms of spacing, where $d$ is the dimensionality of the mesh. 

A local relaxation procedure is undertaken in an attempt to determine a quasi-optimal update. A series of local iterates are considered, setting $\Delta_{i}^{m} = (\frac{1}{2})^{m}$ for $m \leq 5$; terminating as soon an improvement in the worst-case incident-quality metric $\mathcal{Q}_{i}^{\mathcal{T}}(\mathbf{X})$ is achieved. This constrained update strategy is designed to achieve monotonic behaviour; guaranteed to only enact vertex updates that improve the worst-case incident grid quality metrics. In cases where no improvement is possible, an alternative gradient-ascent type strategy is pursued.

\subsubsection{Direct gradient-ascent}

\medskip

While the OWT-type iteration is effective in improving the quality and mesh-spacing conformance of a grid on average, it is not guaranteed to improve the metrics in all cases. Therefore, an additional gradient-ascent type strategy is pursued \cite{freitag1997tetrahedral,klingner2008aggressive,engwirda2017jigsaw}, in which vertex positions are adjusted based on the local gradients of incident element-quality functions $\mathcal{Q}^{\mathcal{T}}(\mathbf{X})$, following a procedure similar to that developed for weight-selection in Section~\ref{section_dual_optim}. Specifically, a given vertex $\mathbf{x}_{i} \in \mathcal{T}(\mathbf{X},W)$ is repositioned along a local \textit{steepest-ascent} vector, chosen to improve the quality of the worst incident element
\begin{equation}
\label{eqn_tria_optim_update}
\mathbf{x}_{i}^{n+1} = \mathbf{x}_{i}^{n} + \Delta_{i}^{m}\, {\reviewerB \mathbf{v}_{i}^{n}}\,,
\end{equation}
where
\begin{equation}
\mathbf{v}_{i}^{n} = \frac{\partial}{\partial \mathbf{x}_{i}}\, \mathcal{Q}^\mathcal{T}_{\,j}(\mathbf{X})
\quad \text{and}\, \quad
j = \operatorname{argmin}_{k} \mathcal{Q}^\mathcal{T}_{\,k}(\mathbf{X}) \,\,\forall \text{ adj. } \tau_{k}\in\mathcal{T}(\mathbf{X},W)\,.
\end{equation}
Consistent with the steepest-ascent strategy outlined in Section~\ref{section_dual_optim}, the index $k$ is taken as a loop over the primal triangles $\tau_{k} \in\mathcal{T}(\mathbf{X},W)$ incident to the vertex $\mathbf{x}_{i}$, $\Delta_{i}^{m} \in \mathbb{R}^{+}$ is a scalar step-length, and the ascent vector $\mathbf{v}_{i}^{n}$ is taken as the gradient of the \textit{worst} incident quality metric. Such a strategy is designed to provide updates in a \textit{worst-first} manner as a result.

A local line search is performed via a simple bisection strategy, iteratively reducing the step-length such that $\Delta_{i}^{m} = (\frac{1}{2})^{m}\, \Delta\bar{\mathbf{x}}$, where $m$ is the local line search iterate. The initial guess $\Delta\bar{\mathbf{x}}$ is computed by considering a first-order Taylor expansion in local grid quality metrics
\begin{equation}
\label{eqn_dual_optim_step}
\mathcal{Q}^\mathcal{T}_{\,j}(\mathbf{X}) + \Delta\bar{\mathbf{x}}\, \frac{\partial}{\partial \mathbf{x}_{i}}\, \mathcal{Q}^\mathcal{T}_{\,j}(\mathbf{X}) \leq \bar{\mathcal{Q}}^\mathcal{T}_{\,k}(\mathbf{X})\,.
\end{equation} 
Here, $\bar{\mathcal{Q}}^\mathcal{T}_{\,k}(\mathbf{X})$ is a taken as a mean quality value over the local incident set $\tau_{k} \in \mathcal{T}(\mathbf{X},W)$, with the corresponding $\Delta\bar{\mathbf{x}}$ an estimate of the perturbation required to improve the worst metric until it is equal to this local mean measure. A limited line search is employed, testing iterations $m \leq 5$ until a successful step is found. If no such improvement is identified, the vertex is left unchanged, with $\mathbf{x}_{i}^{n+1} = \mathbf{x}_{i}^{n}$. Consistent with previous discussions, such a procedure is guaranteed to monotonically improve the worst-case grid quality metric in the incident set, constituting a \textit{quality-preserving} strategy.

\subsection{Incremental updates to mesh topology}

\begin{figure*}[t]
  \centering  
  \includegraphics[width=.8625\textwidth]{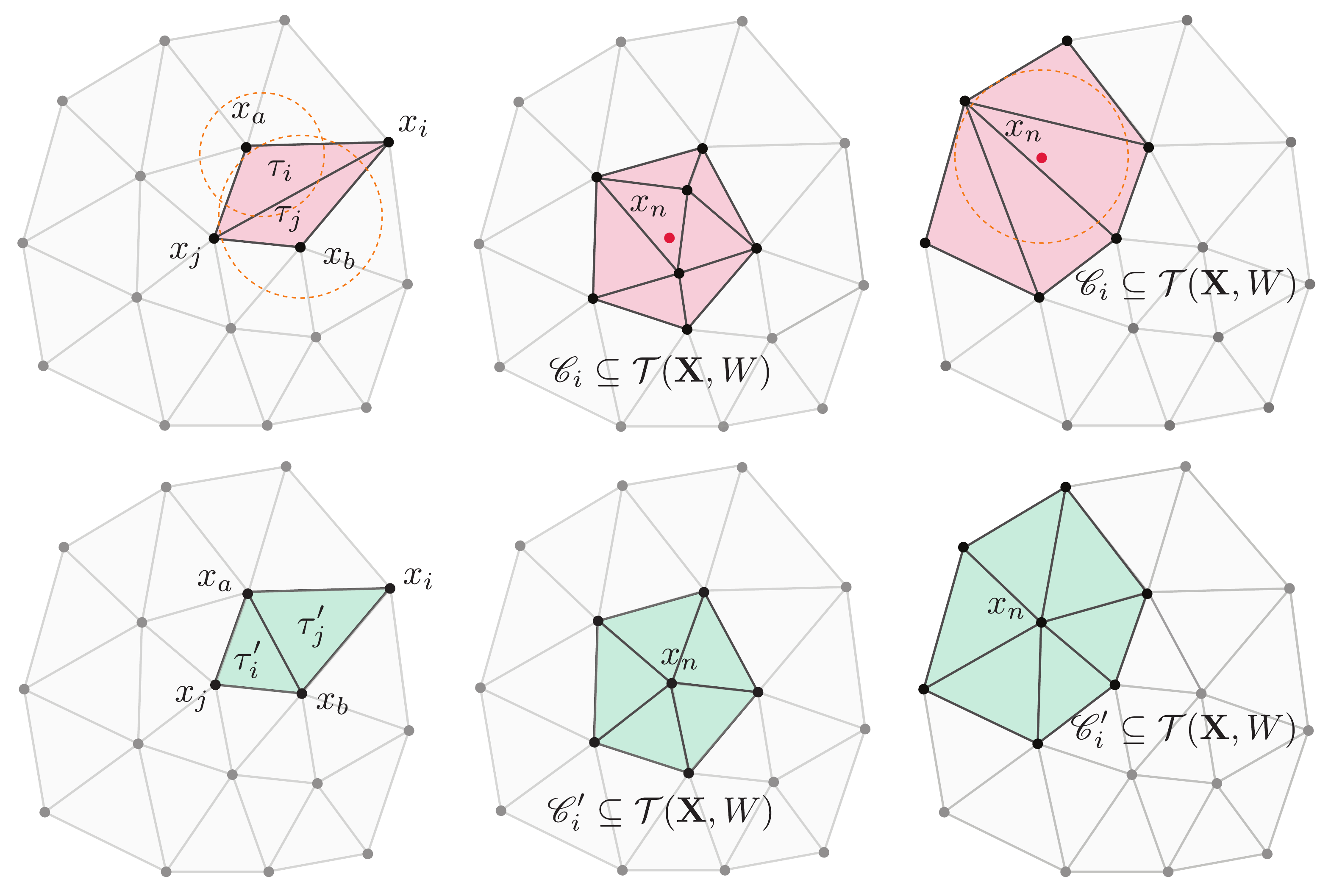}
  
  \caption{Topological operations for primal grid optimisation, showing (left) an edge-flip, (middle) an edge-collapse, and (right) an edge-refinement operation. Grid configurations before and after each update are shown in the upper and lower panels, respectively.}
  
  \label{figure_topo_flip}
\end{figure*}

\medskip

Following perturbations to the geometry of the primal and/or dual grid, adjustments must be made to the underlying mesh topology, such that the triangulation remains a valid regular tessellation and the dual its orthogonal power diagram. While it is possible to simply re-compute the full weighted triangulation after each adjustment is made, such an approach would impose a significant computational burden, especially when considering that a majority of updates involve small perturbations to the vertex positions or weights; unlikely to require large-scale changes to the grid topology. In this work, an alternative strategy is pursued, based on local element-wise transformations, known as topological \textit{flips}.

Given a pair of adjacent triangles $\{\tau_{i},\tau_{j}\} \in \mathcal{T}(\mathbf{X},W)$ a local re-triangulation can be achieved by \textit{flipping} the local connectivity about the shared edge $\{x_{i},x_{j}\}$, forming a new edge between the opposing vertices $\{x_{a},x_{b}\}$. This operation replaces the existing triangle pair $\{\tau_{i},\tau_{j}\}$ with a new set $\{\tau_{i}',\tau_{j}'\}$. See Figure~\ref{figure_topo_flip}a for details. In the present work, the iterative application of such edge-flipping operations is used to incrementally transform the topology of the primal triangulation; ensuring that the local weighted tessellation criterion is satisfied throughout the optimisation process. Specifically, given a general triangulation $\mathcal{G}(\mathbf{X},W)$, possibly violating the local weighting criterion, a cascade of edge-flips are employed to transform $\mathcal{G}(\mathbf{X},W)$ into a valid regular triangulation $\mathcal{T}(\mathbf{X},W)$. For each adjacent pair $\{\tau_{i},\tau_{j}\} \in \mathcal{G}(\mathbf{X},W)$, a flip is enacted if a local violation of the weighted criterion is detected. New triangles created by successful edge flips are subsequently re-examined until no further modifications are required. This approach is a generalisation of the standard incremental edge-flipping algorithms for Delaunay triangulations, previously presented by various authors, including, for example, Lawson \cite{lawson77flips}.

Given a pair of adjacent triangles $\{\tau_{i},\tau_{j}\} \in \mathcal{T}(\mathbf{X},W)$, the local weighted criterion is violated when the power distance $\pi_{a}(\mathbf{o}_{j})$ and $\pi_{b}(\mathbf{o}_{i})$ between the staggered vertices $\{(x_{a},w_{a}),(x_{b},w_{b})\}$ and the opposing orthocentres $\{\mathbf{o}_{j},\mathbf{o}_{i}\}$ is less than the power-radii of the orthogonal balls associated with $\tau_{i}$ and $\tau_{j}$. Such a constraint is equivalent to requiring that adjacent cells in the dual power diagram be non-overlapping. To prevent issues with exact floating-point comparisons, a tolerance-based comparison strategy is employed. Specifically, both opposing vertices are required to penetrate their opposing orthoball by a squared distance $\epsilon$ before a flip is enacted, with $\epsilon = \max(r_{i}^{2},r_{j}^{2})\,\bar{\epsilon}$ and $\bar{\epsilon} = 1 \times 10^{-10}$ in the current double-precision implementation. See Figure~\ref{figure_topo_flip}a for details.

\subsection{Refinement \& edge collapse}

\medskip

In addition to updates to the bulk geometry and topology of the grid, mesh quality and mesh spacing conformance can often be further improved through the addition and/or removal of vertices. In the present work, such behaviour is achieved through a set of \textit{edge-refinement} and \textit{edge-collapse} operations, generalising the methodology presented by the author in \cite{engwirda2017jigsaw} to handle general weighted primal-dual tessellations.

Given an edge $e_{k}$ in the primal tessellation $\mathcal{T}(\mathbf{X},W)$, a \textit{collapse} operation is achieved by \textit{merging} the two vertices $\{x_{i},\,x_{j}\} \in e_{k}$ to a local mean position $\mathbf{x}_{m}$ and re-triangulating the local cavity $\mathcal{C}_{k} \subseteq \mathcal{T}(\mathbf{X},W)$ incident to the edge. In the present work, vertices are merged to an average of adjacent element orthocentre coordinates, such that $\mathbf{x}_{m} = \frac{1}{|\mathcal{C}_{k}|}\sum{\mathbf{o}_{i}}$ for all local triangles $\tau_{i} \in \mathcal{C}_{k}$. While such an approach is slightly more computationally intensive than simpler strategies based on collapsing vertices to primal edge midpoints, the use of averaged orthocentres has been found to be substantially more effective in practice. See Figure~\ref{figure_topo_flip}b for details.

Given an edge $e_{k}$ in the primal tessellation $\mathcal{T}(\mathbf{X},W)$, a \textit{refinement} operation is achieved by inserting a new vertex $x_{n}$, positioned at the centre of the orthoball associated with the lower quality adjacent triangle $\tau_{i} \in \mathcal{T}(\mathbf{X},W)$. Insertion of the new vertex $x_{n}$ induces a re-triangulation of the local cavity $\mathcal{C}_{k} \in \mathcal{T}(\mathbf{X},W)$; constructed by expanding about $x_{n}$ in a local greedy fashion. Starting from the initial cavity $\mathcal{C}_{k} = \{\tau_{i},\tau_{j}\}$, where $\{\tau_{i},\tau_{j}\} \in \mathcal{T}(\mathbf{X},W)$ are the triangles adjacent to the edge $e_{k}$, additional elements are added to the cavity $\mathcal{C}_{k}$ in a breadth-first manner, with a new, unvisited neighbour $\tau_{k}$ added if doing so will improve the worst-case grid-quality metric $\mathcal{Q}^{\mathcal{T}}(\mathbf{X})$ of the re-triangulated configuration. The final cavity $\mathcal{C}_{k}$ is therefore a locally optimal stencil for the insertion of $x_{n}$. In practice, this iterative deepening of $\mathcal{C}_{k}$ typically convergences in one or two iterations, and can be implemented efficiently using local adjacency information. See Figure~\ref{figure_topo_flip}c for details. 

Both the edge-collapse and edge-refinement operations are designed to achieve an improvement in worst-case primal grid quality metrics. A sweep over primal mesh edges is performed, with both a collapse and refinement operation is first \textit{simulated} for each edge. Given an edge $e_{k}$, a subsequent collapse or refinement operation is then \textit{accepted} if the re-triangulated cavity ${\mathcal{C}_{k}}'$ is of improved quality, such that $\min \mathcal{Q}_{k'}^{\mathcal{T}}(\mathbf{X}) \geq \min \mathcal{Q}_{k}^{\mathcal{T}}(\mathbf{X}) + \bar{\epsilon}$, where $\bar{\epsilon} = 1 \times 10^{-8}$ is a quality improvement threshold, and the indices $k$ and $k'$ iterate over the local cavities, considering all $\tau_{k} \in \mathcal{C}_{k}$ and $\tau_{k}' \in \mathcal{C}_{k}'$.

\subsection{A coupled optimisation schedule}

\begin{algorithm*}[t]
\caption{A Coupled Primal-dual Optimisation Strategy}
\label{algorithm_primal_dual_optim}
\centering

 {
 \reviewerB

 \smallskip

 \begin{algorithmic}[1]
 
 \Function{OptimisePrimalDual}{$\mathbf{X},W,\mathcal{T},\mathcal{D}$}

  \smallskip

  \State // \textit{Optimise a given primal-dual pair $(\mathcal{T},\mathcal{D})$, employing}
  \State // \textit{a combination of geometrical, topological and weight-}
  \State // \textit{based operations.}

  \smallskip

  \For{$n = 1 \text{ to } N$} \tabto{.5\textwidth} // \textit{coupled outer iterations}
  
   \smallskip
  
   \For{$m = 1 \text{ to } M$} \tabto{.5\textwidth} // \textit{vertex + weight updates}
   
    \smallskip
   
    \State $(\mathbf{X},W)^{m} \leftarrow \textsc{radnperm}\left((\mathbf{X},W)^{n}\right)$
   
    \smallskip
   
    \For{$ \text{all } x_{i} \in \mathcal{T}(\mathbf{X},W)$} \tabto{.5\textwidth} // \textit{vertex updates}
 
     \State Find update direction $\mathbf{v}_{i} = \textsc{grad-x}((\mathbf{X},W)^{m+1})$ 
 
     \State Perform line search $\mathbf{x}_{i}^{*} \leftarrow \mathbf{x}_{i}^{m}\, + \Delta_{i}\,\mathbf{v}_{i}$
 
     \State Monotonic updates $\mathbf{x}_{i}^{m+1} \leftarrow \mathbf{x}_{i}^{*}, \text{ iff }\, \textsc{better}((\mathbf{X},W)^{*})$
 
    \EndFor
    
    \smallskip
    
    \For{$ \text{all } w_{i} \in \mathcal{T}(\mathbf{X},W)$} \tabto{.5\textwidth} // \textit{weight updates}
 
     \State Find update direction $v_{i} = \textsc{grad-w}((\mathbf{X},W)^{m+1})$ 
 
     \State Perform line search $w_{i}^{*} \leftarrow w_{i}^{m}\, + \Delta_{i}\, v_{i}$
 
     \State Monotonic updates $w_{i}^{m+1} \leftarrow w_{i}^{*}, \text{ iff }\, \textsc{better}((\mathbf{X},W)^{*})$
 
    \EndFor
   
    \smallskip
   
   \EndFor
  
   \smallskip
  
   \State // \textit{Update grid topology}
  
   \smallskip
  
   \State $\mathcal{T}(\mathbf{X},W)^{n+1} \leftarrow \textsc{fliptopology}\left(\mathcal{T}(\mathbf{X},W)^{n}\right)$
   
   \smallskip
   
   \State // \textit{Refine/collapse edge}
   
   \smallskip
   
   \State $\mathcal{T}(\mathbf{X},W)^{n+1} \leftarrow \textsc{pruneedges}\:\left(\mathcal{T}(\mathbf{X},W)^{n+1}\right)$
   
   \State $\mathcal{T}(\mathbf{X},W)^{n+1} \leftarrow \textsc{refineedges}  \left(\mathcal{T}(\mathbf{X},W)^{n+1}\right)$
  
   \smallskip
  
  \EndFor

  \smallskip

  \State \Return optimised primal-dual complexes $\mathcal{T}(\mathbf{X},W)$ and $\mathcal{D}(\mathbf{X},W)$

  \smallskip

 \EndFunction
 
 \end{algorithmic}

 \smallskip

  }

\end{algorithm*}

\medskip

The full primal-dual grid optimisation procedure is realised as a combination of the various geometrical and topological operations described previously; organised into a particular iterative optimisation schedule. See Algorithm~\ref{algorithm_primal_dual_optim} for details. Each outer iteration consists of a fixed set of operations: eight sweeps to update vertex positions and weights, an iterative edge-flipping scan to restore the local `weighted' triangulation criterion, and, finally, a single pass of edge refinement/collapse operations. {\reviewerC It was found that combining multiple vertex and weight updates with a single stage of topological transformation led to enhanced results in practice, allowing substantial improvement to the geometry of the grid to be achieved before making discrete changes to its topology.} In this study, a maximum of sixteen outer iterations were employed. {\reviewerC In typical cases, it was found that additional outer iterations did not further improve the worst-case grid quality metrics.} Each vertex- and weight-update pass is implemented as a composite operation, with the variational OWT-type technique supplemented with local gradient-ascent iterations as required. Given a vertex $\mathbf{x}_{i}$ in the primal mesh, an OWT update is always attempted first, with a subsequent gradient-ascent step employed in cases where local grid-quality metrics are sufficiently improved y the initial OWT step. Updates to the vertices $\mathbf{X}$ and weights $W$ occur sequentially, with a single, linear sweep over the grid vertices followed by a single pass over the mesh weights. Each vertex- and weight-update pass is arranged to follow a weakly stochastic, Gauss-Seidel type philosophy; with vertices and weights visited in a randomised order, and updated position and weight values leveraged in subsequent calculations soon as they are available. The coupled optimisation schedule employed here is not based on a particular theoretical derivation, but is simply a set of heuristic choices that have proven to be effective in practice, building on previous primal mesh optimisation techniques \cite{freitag1997tetrahedral,klingner2008aggressive}. A similar mesh improvement strategy, combining a series of geometrical and topological operations, was previously pursued by the author in \cite{engwirda2017jigsaw} for the optimisation of conventional Delaunay-Voronoi structures.

\section{Experimental results}
\label{section_results}

The performance of the primal-dual optimisation algorithms presented in Sections~\ref{section_dual_optim} and \ref{section_primal_dual_optim} was investigated experimentally, with the methods employed to generate a set of optimised Regular-Power tessellations for a series of benchmark problems. A pair of test-cases appropriate for regional ocean modelling applications were investigated. All schemes have been implemented in C++. The full algorithm has been implemented as part of the JIGSAW meshing package, and is currently available online \cite{JIGSAW-GEO} or by request from the author. All tests were completed on a Linux platform, using a single core of an Intel i7 processor. Visualisation and post-processing was completed using MATLAB.

\subsection{Preliminaries}

\begin{figure*}[t]
\centering
\includegraphics[width=.3875\textwidth]{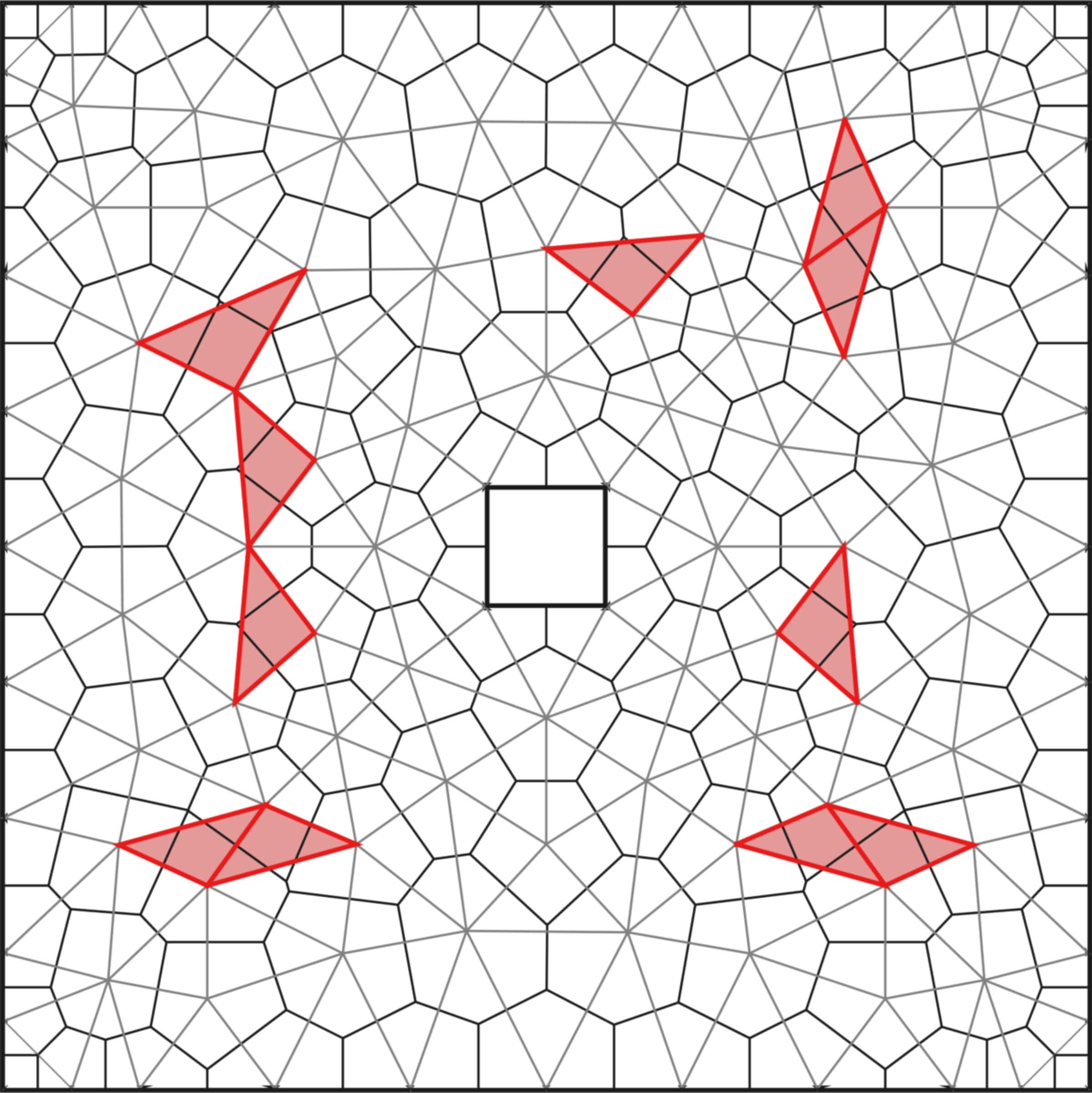}\qquad
\includegraphics[width=.3875\textwidth]{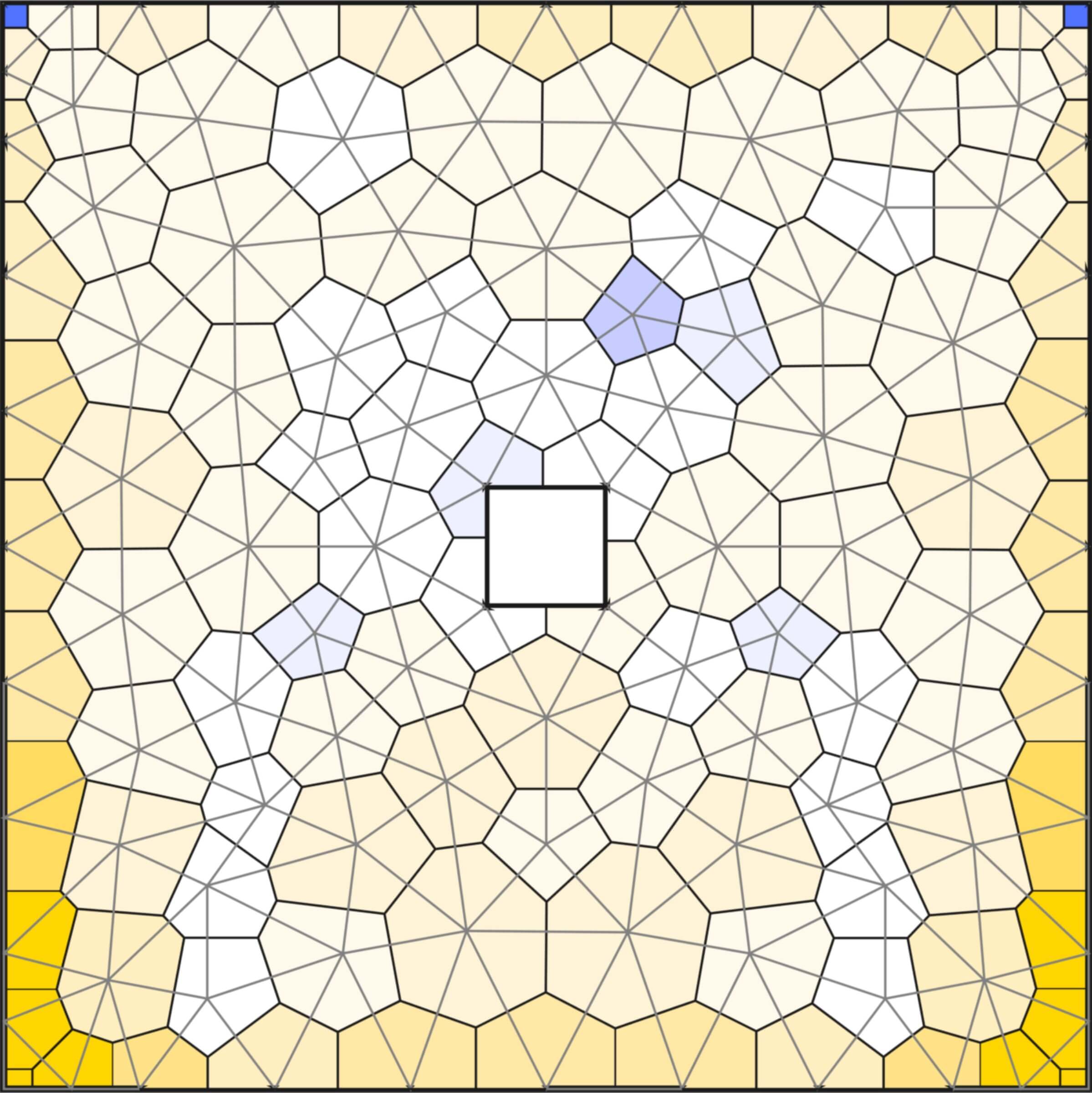}

\captionof{figure}{Primal-dual grids for the Box test-case, showing: (i) the initial grid, and (ii) a generalised dual grid generated using the weight-selection algorithm. Both the primal triangulation $\mathcal{T}(\mathbf{X},W)$ and the dual power diagram $\mathcal{D}(\mathbf{X},W)$ are shown here. Dual cells are coloured by their \textit{relative power}, with blue tones indicating negative values, orange tones positive values, and white tones associated with values approaching zero. Triangles highlighted in red violate the well-centred constraint, and do not contain their own dual vertices. Note that this is a \textit{dual-only} example; no optimisation of the primal grid is conducted.}

\label{fig_box_mesh}

\bigskip

\includegraphics[width=.3875\textwidth]{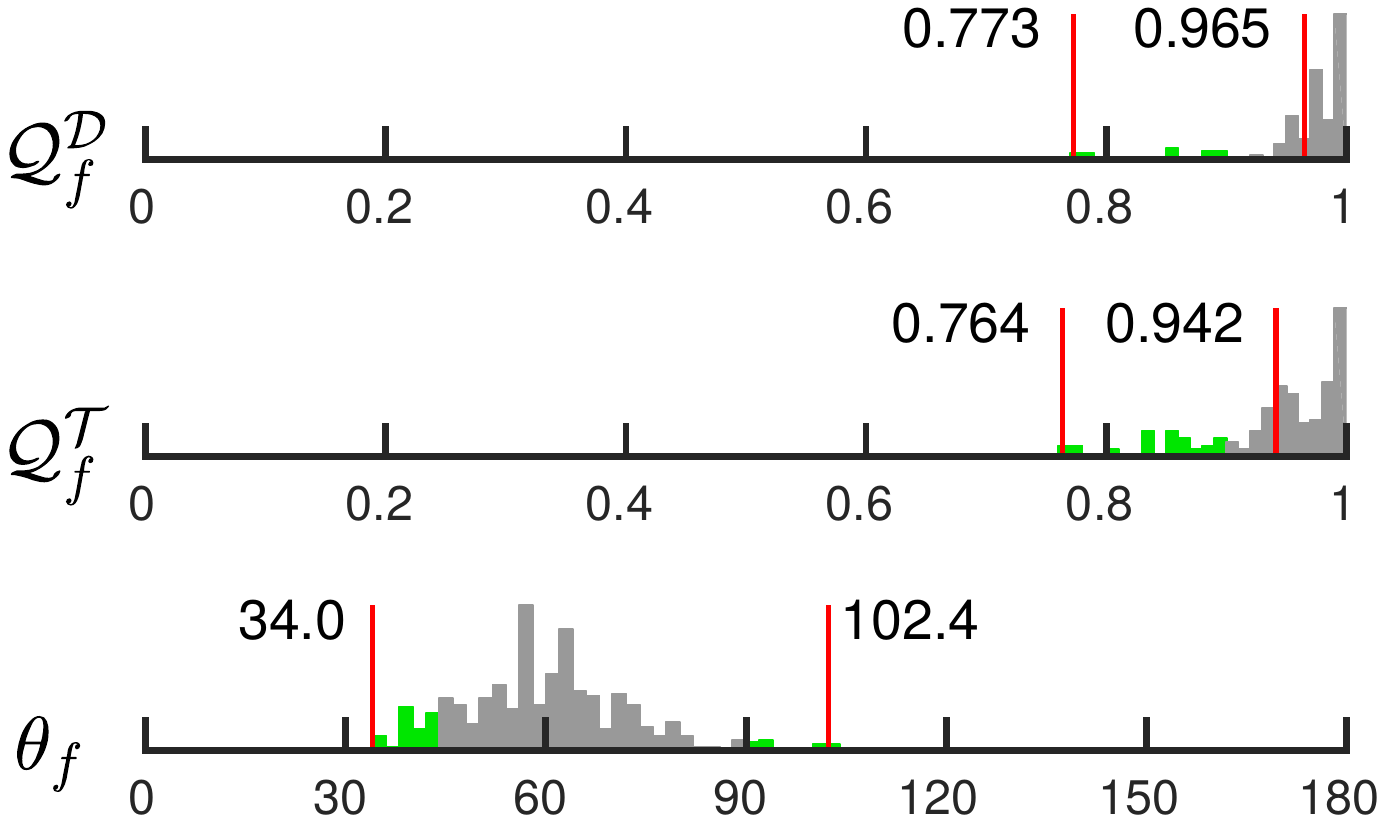}\qquad
\includegraphics[width=.3875\textwidth]{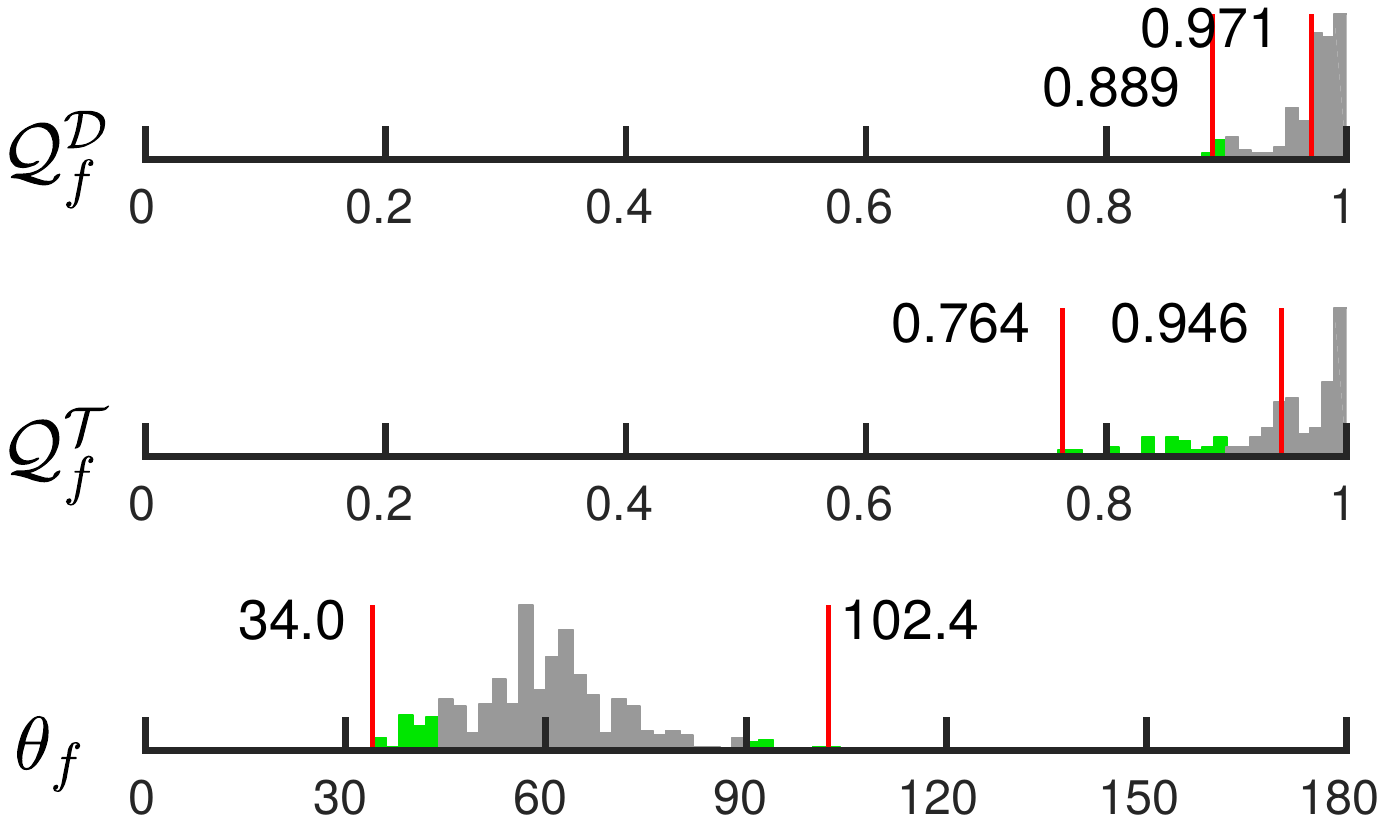}

\captionof{figure}{Histograms of primal-dual quality metrics for the Box test-case, showing data for: (a) the initial grid, and (b) a generalised dual grid generated using the weight-selection algorithm. Note that this is a \textit{dual-only} example; no optimisation of the primal grid is conducted..}

\label{fig_box_cost}
\end{figure*}

\medskip

Three benchmark problems were considered; the first designed to assess the performance of the weight selection strategy described in Section~\ref{section_dual_optim} and the second and third designed to test the coupled primal-dual strategy developed in Section~\ref{section_primal_dual_optim}. In the first example, a generalised dual grid is generated for a given static primal tessellation, using the locally-optimal weight-selection strategy to improve the quality of the dual grid geometry while keeping the underlying primal mesh unchanged. In the second and third examples, the fully coupled optimisation problem is considered; employing the full suite of optimisation techniques to select optimal vertex positions, vertex weights and mesh topology. In all cases, the initial primal mesh consists of a high-quality conforming Delaunay triangulation; generated using the JIGSAW library. All problems are initialised with $W^{\,n=0}=\{0\}$, implying $\mathcal{T}(\mathbf{X},W) = \operatorname{Del}(\mathbf{X})$ and $\mathcal{D}(\mathbf{X},W) = \operatorname{Vor}(\mathbf{X})$. The primal and dual grids associated with each test-case are shown in Figures~\ref{fig_box_mesh}, \ref{fig_australia_mesh} and \ref{fig_antarctica_mesh}, with associated distributions of grid-quality metrics presented in Figures~\ref{fig_box_cost} and \ref{fig_grids_cost}. Grid quality statistics and computational performance is summarised in Table~\ref{table_summary}.

For all test problems, detailed statistics on primal and dual cell quality are presented, including histograms of the primal and dual quality metrics $\mathcal{Q}^{\mathcal{T}}(\mathbf{X})$ and $\mathcal{Q}^{\mathcal{D}}(\mathbf{X},W)$ defined in (\ref{eqn_dual_quality}) and (\ref{eqn_tria_quality}). The distribution of angles in the primal mesh $\theta_{\tau}$ {\reviewerC as well as the \textit{relative-length} metric $h_{r}(e_{i}) = \nicefrac{\|\mathbf{e}_{i}\|}{\bar{h}_{e}}$ are also presented, where $h_{r}(e_{i})$ represents conformance to the imposed mesh-spacing function $\bar{h}(\mathbf{x})$ sampled at each primal edge $e_{i} \in \mathcal{T}(\mathbf{X},W)$.} Histograms highlight the minimum, maximum, and mean values of the relevant distributions as appropriate. {\reviewerB Additional measures of \textit{spread} in the distribution of angles and mesh-spacing conformance are also provided, defined as the \textit{Mean Absolute Deviation} in $\theta_{\tau}$ and $h_{r}$, where
\begin{gather}
\bar{\sigma}_{\theta} = \frac{1}{n_{\theta}}\sum_{j=1}^{n_{\theta}}|\bar{\theta}_{\tau} - \theta(\tau_{j})|\,,
\qquad
\bar{\sigma}_{h} = \frac{1}{n_{h}}\sum_{j=1}^{n_{h}}|\bar{h}_{r} - h_{r}(\mathbf{x}_{j})|\,,
\end{gather}
where $\bar{\theta}_{\tau}$ and $\bar{h}_{r}$ denote the means of their respective distributions. Smaller values of $\bar{\sigma}_{\theta}$ and $\bar{\sigma}_{h}$ indicate distributions that are more tightly clustered about their mean values, consistent with meshes of higher average quality.} A measure of the distribution of weights in each grid is also provided, represented by what is here called the \textit{relative power} associated with each cell in the dual tessellation
\begin{equation}
W_{r} = \frac{w_{i}}{A_{i}}\,.
\end{equation}
Here, $w_{i}$ is the scalar weight assigned to the primal vertex associated with a given dual cell $D_{i} \in \mathcal{D}(\mathbf{X},W)$ and $A_{i}$ is the area of $D_{i}$. Noting that the weights $w_{i}$ can be interpreted as the squared-radii of circles positioned at the primal vertices \cite{aurenhammer1987power}, the relative power is a measure of the \textit{relative} magnitude of the weights applied throughout a mesh. $W_{r}$ is a non-dimensional metric; allowing for a comparison of weights applied throughout graded tessellations where cells are of varying sizes. Contours of $W_{r}$ are presented for each test-case considered in the following sections; illustrating the distribution of weights generated by the new primal-dual optimisation scheme.

\subsection{Standalone dual grid optimisation}

\medskip

In the `Box' test-case shown in Figure~\ref{fig_box_mesh}, the performance of the weight-selection algorithm developed in Section~\ref{section_dual_optim} is investigated; seeking to generate an optimised Power diagram given a static primal tessellation. This test problem was designed as a simple verification case and consists of a coarse Delaunay triangulation of a concentric square geometry. The Delaunay-refinement algorithm available in the JIGSAW package \cite{engwirda2017jigsaw,JIGSAW-GEO} was used to provide the initial configuration; defined as a \textit{quality} Delaunay tessellation, such that all angles in the primal elements exceed $30^\circ$. The associated unperturbed dual is the conventional Voronoi diagram. A set of primal-dual pairs are shown in Figure~\ref{fig_box_mesh}, illustrating the effect of weight optimisation on the geometry of the grid. Dual cells are coloured by their relative power $W_{r}$, showing the distribution of weights throughout the grid. Poorly staggered elements are also highlighted, identifying primal triangles that do not contain their associated dual vertices. Overall mesh quality is quantified in Figure~\ref{fig_box_cost}, where histograms of primal and dual grid quality metrics illustrate the distribution of $\mathcal{Q}^{\mathcal{D}}(\mathbf{X},W)$, $\mathcal{Q}^{\mathcal{T}}(\mathbf{X})$ and $\theta(\tau)$ before and after the application of the weight-selection scheme.

\subsubsection{Discussions}

\medskip

A comparison of the primal-dual pairs shown in Figure~\ref{fig_box_mesh} reveals the effectiveness of the proposed weight-selection procedure, demonstrating that a generalised dual structure can be generated that is both more centroidal and more well-centred than the standard Delaunay-Voronoi configuration. In the initial state, 11 triangles are poorly staggered, with their associated dual vertices lying exterior to the elements themselves. The optimised primal-dual tessellation contains no poorly staggered triangles, with dual vertices shifted toward their associated triangle centroids through the selection of an appropriate set of vertex weights. Improvements to the structure of the dual are confirmed through an analysis of the grid quality metrics presented in Figure~\ref{fig_box_cost}, confirming that both the minimum and mean values of $\mathcal{Q}^{\mathcal{D}}(\mathbf{X},W)$ are improved in the generalised weighted configuration. In this case, $\mathcal{Q}_{\text{min}}^{D}: 0.773 \rightarrow 0.889$ and $\bar{\mathcal{Q}}^{D}: 0.965 \rightarrow 0.971$, respectively. Note that a slight change in the primal metric $\mathcal{Q}^{\mathcal{T}}(\mathbf{X})$ is also recorded; due to differences in the topology of the associated Delaunay and Regular triangulations. Though a direct optimisation of the primal grid is not employed in this example, the topology of the triangulation is modified as required; ensuring that the primal-dual pair remains orthogonal. In this case, optimisation of the vertex weights leads to a single edge flip in the Regular triangulation.

\subsection{Coupled primal-dual optimisation}

\begin{table*}
\centering
\caption{Summary of results for the East Australia and Antarctic Peninsula test-cases, enumerating: the size of the primal triangulation $|\mathcal{T}|$, algorithm runtime \textsc{CPU}, minimum and mean primal/dual grid quality metrics $\mathcal{Q}^{\mathcal{T}}$ and $\mathcal{Q}^{\mathcal{D}}$, and the number of poorly staggered triangles $|\tau_{\textsc{bad}}|$. Data is provided for the initial configuration, primal-only optimisation (\textsc{-p}), and coupled primal-dual optimisation (\textsc{-pd}).}

\label{table_summary}

{
\tabulinesep=2pt

\medskip

\begin{tabu} {|l||c|c|c|c|c|c|c|}

\hline

& $|\mathcal{T}|$ & \textsc{cpu} (sec) & 
$\mathcal{Q}_{\text{min}}^{\mathcal{D}}$ & $\bar{\mathcal{Q}}^{\mathcal{D}}$ & 
$\mathcal{Q}_{\text{min}}^{\mathcal{T}}$ & $\bar{\mathcal{Q}}^{\mathcal{T}}$ &
$|\tau_{\textsc{bad}}|$ \\

\hline

\textsc{australia/coral sea}       &  54,017 &   -- & 0.525 & 0.986 & 0.621 & 0.970 & 2073 \\
\textsc{australia/coral sea (-p)}  &      -- & 5.25 & 0.564 & 0.993 & 0.633 & 0.980 &  180 \\
\textsc{australia/coral sea (-pd)} &      -- & 8.54 & 0.857 & 0.997 & 0.633 & 0.983 &    8 \\

\hline

\textsc{antarctic peninsula}       & 145,311 &   -- & 0.450 & 0.983 & 0.594 & 0.965 & 6401 \\
\textsc{antarctic peninsula (-p)}  &      -- & 20.7 & 0.450 & 0.992 & 0.594 & 0.979 &  333 \\
\textsc{antarctic peninsula (-pd)} &      -- & 27.4 & 0.862 & 0.997 & 0.604 & 0.981 &   11 \\

\hline

\end{tabu}}
\end{table*}

\begin{figure*}
\centering
\includegraphics[width=.2850\textwidth]{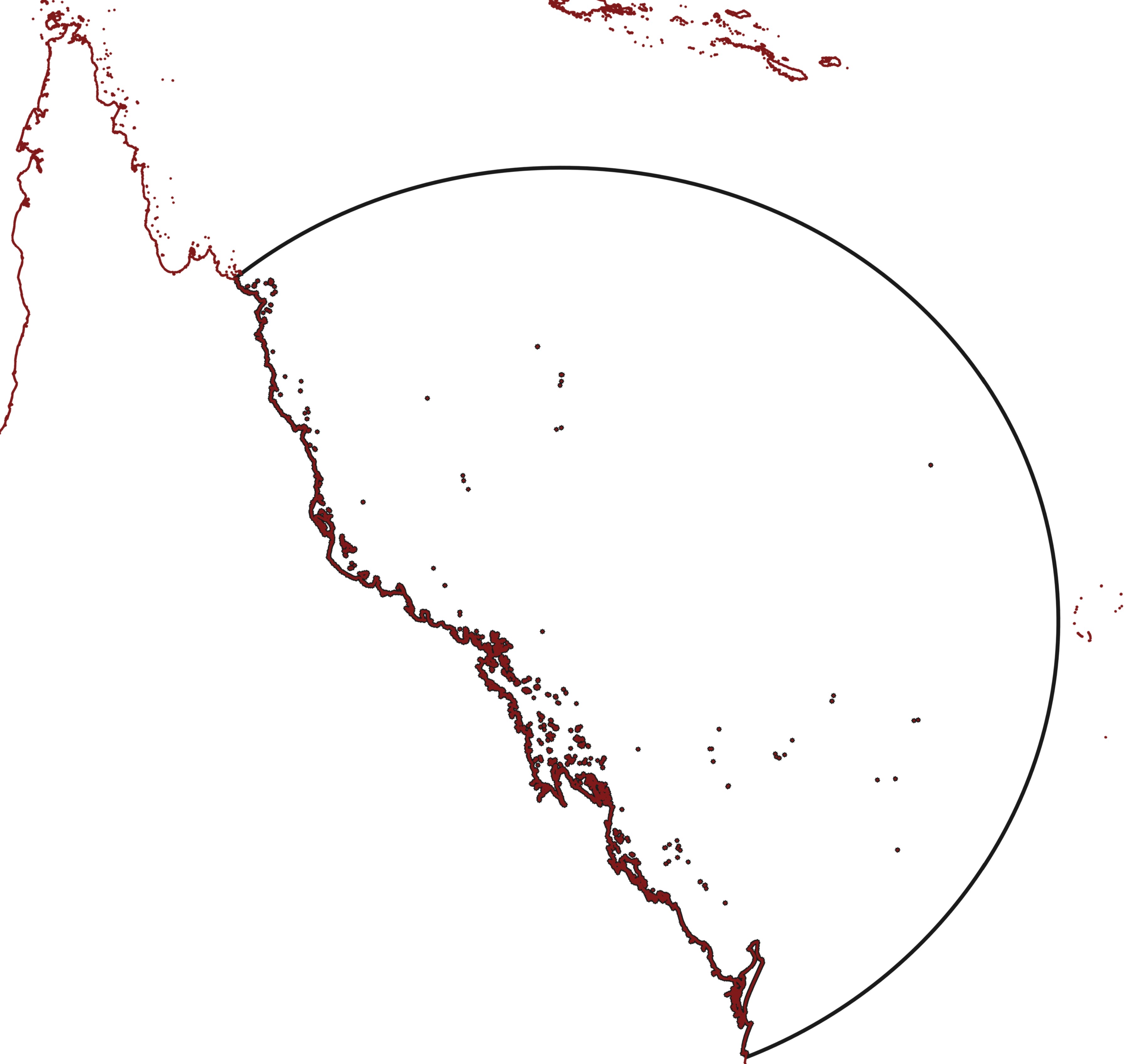}\quad
\includegraphics[width=.2850\textwidth]{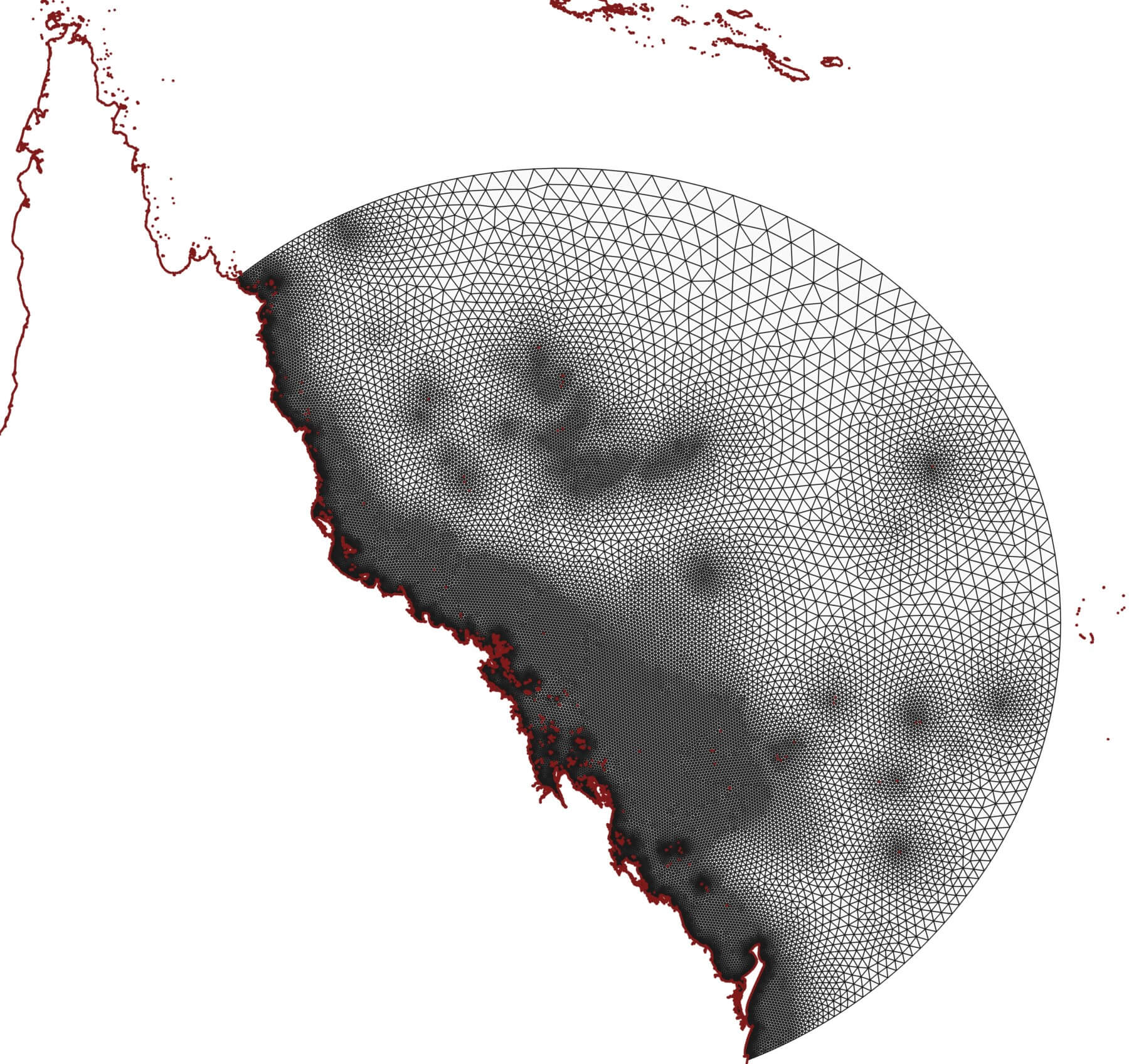}\quad
\includegraphics[width=.2850\textwidth]{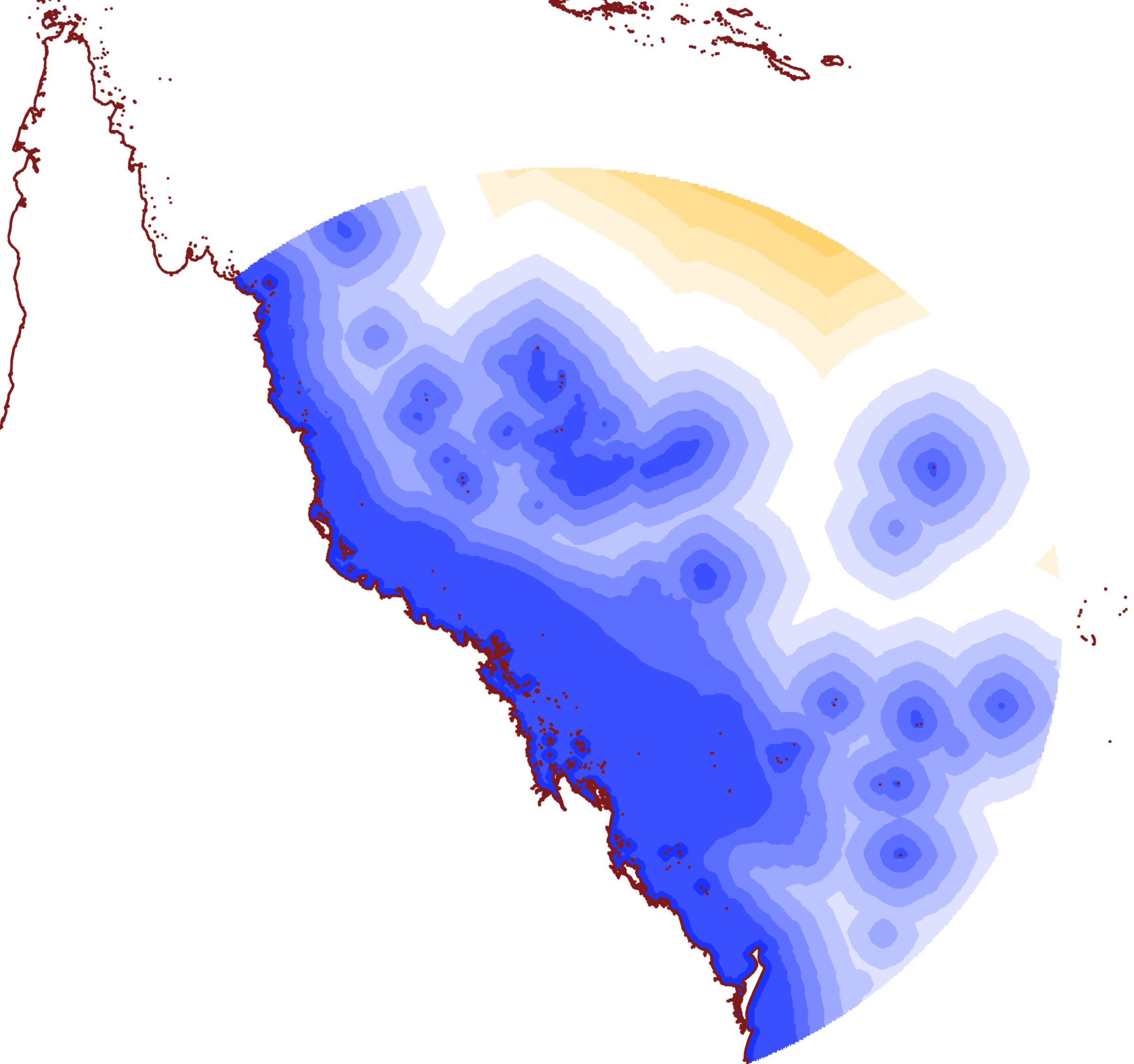}
\\[1ex]
\includegraphics[width=.9250\textwidth]{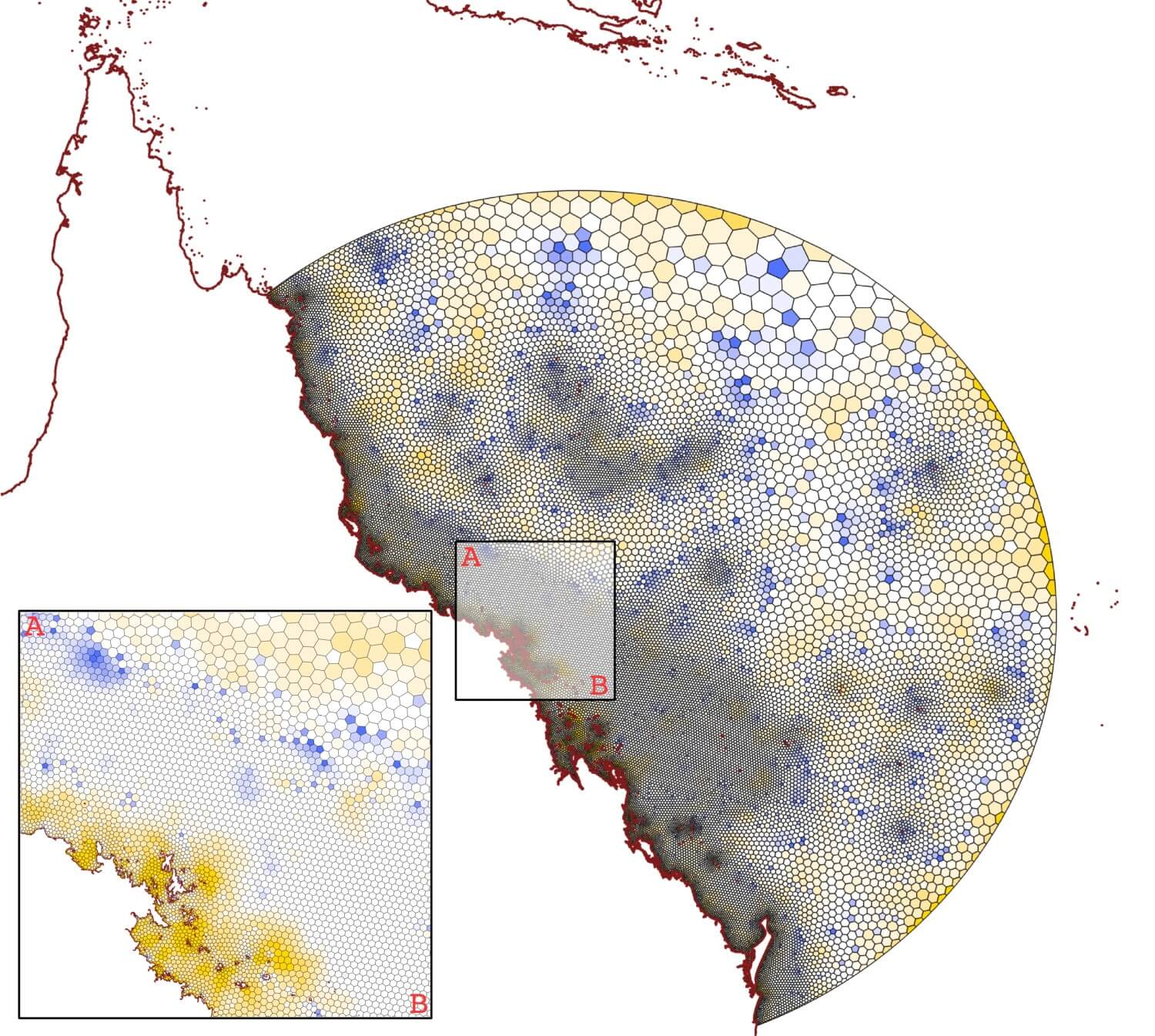}

\caption{An optimised primal-dual grid for the Australian test-case, showing (clockwise from top-left): (a) the coastal geometry and open-boundary definition, (b) the weighted triangulation $\mathcal{T}(\mathbf{X},W)$ (c) contours of the mesh spacing function $\bar{h}(\mathbf{x})$, and (d) the dual power diagram $\mathcal{D}(\mathbf{X},W)$. Here, dual cells are coloured by their \textit{relative power}, with blue tones indicating negative values, orange tones positive values, and white tones associated with values approaching zero.}

\label{fig_australia_mesh}
\end{figure*}

\medskip

\begin{figure*}
\centering
\includegraphics[width=.3000\textwidth]{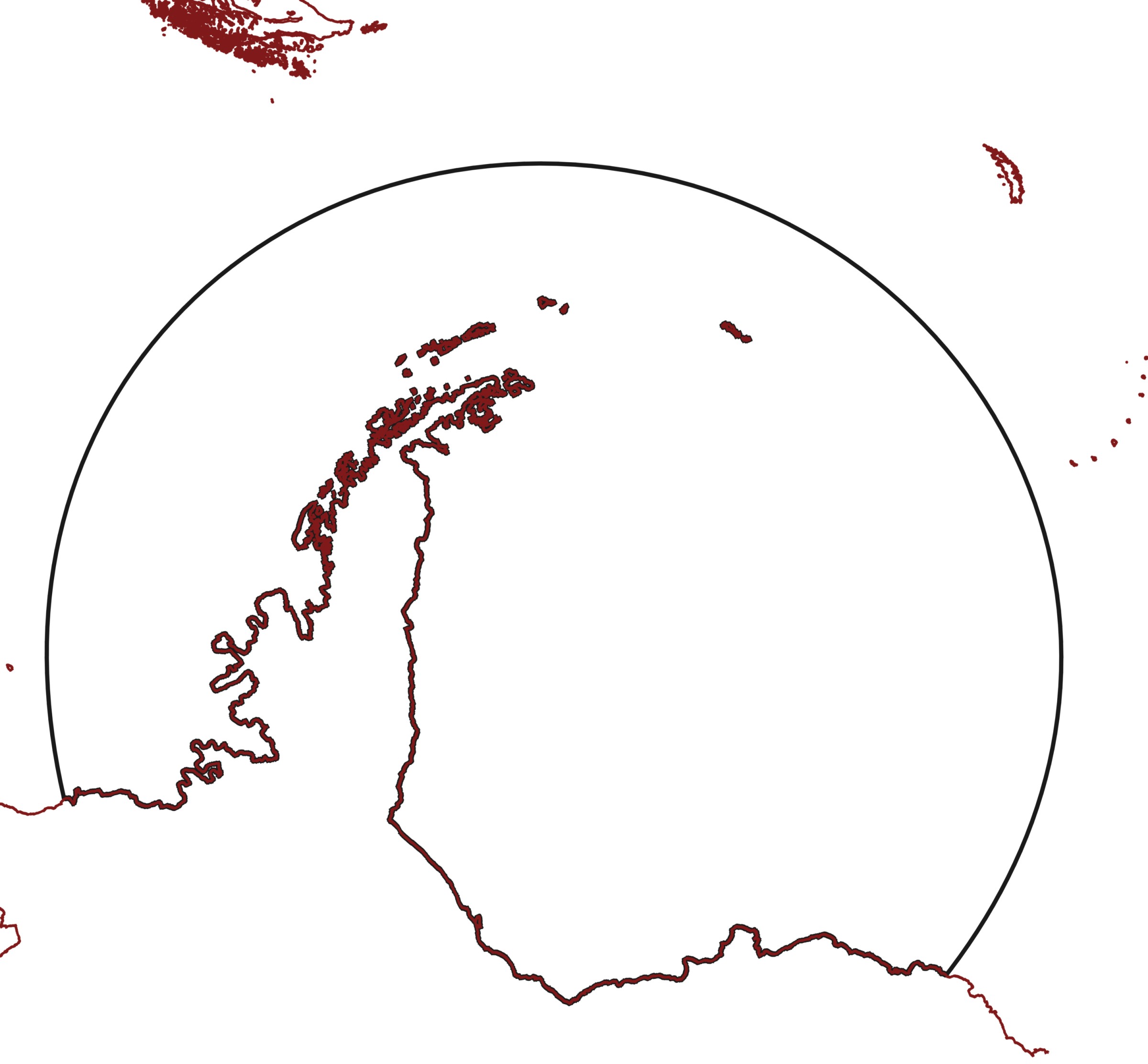}\quad
\includegraphics[width=.3000\textwidth]{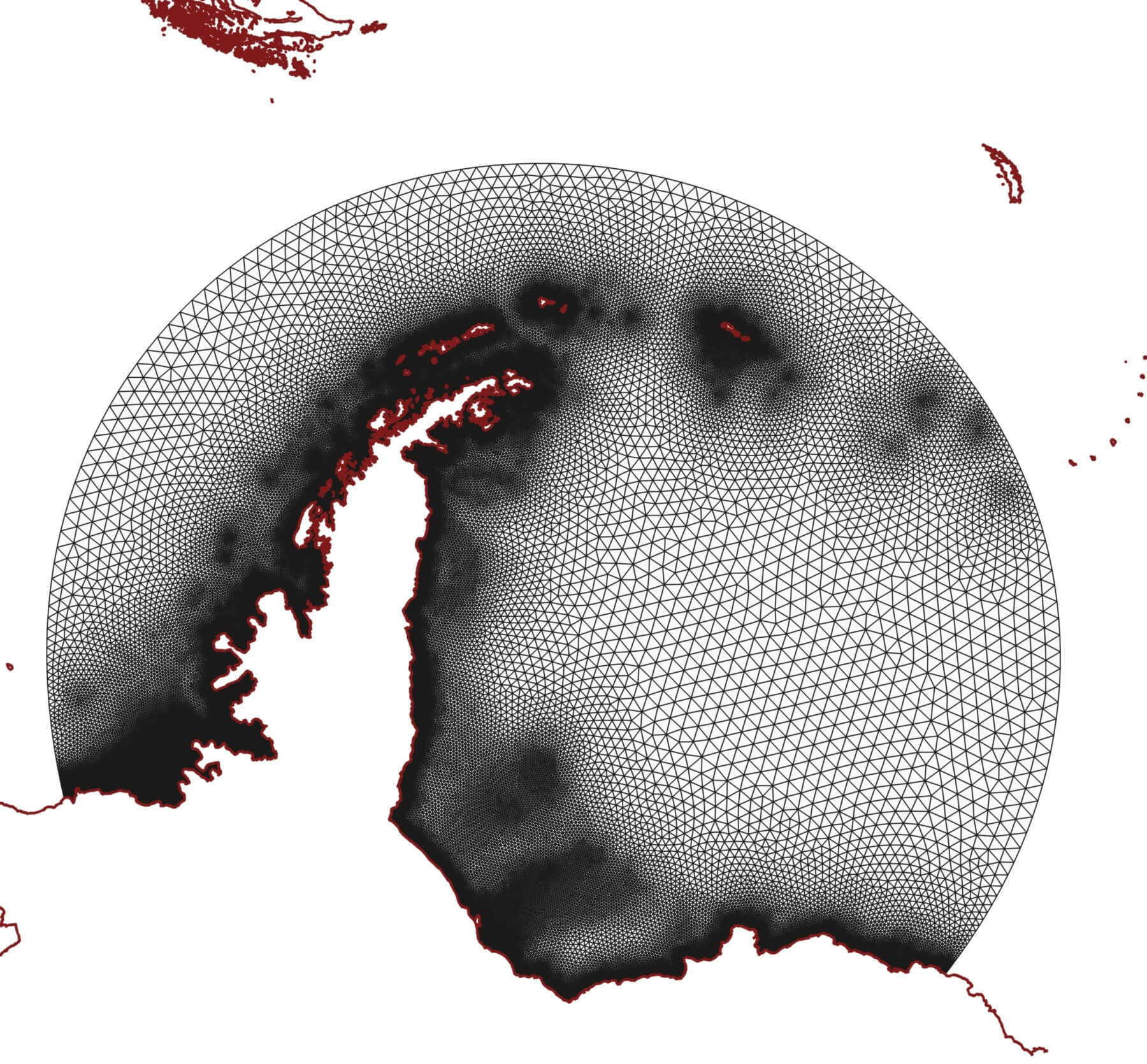}\quad
\includegraphics[width=.3000\textwidth]{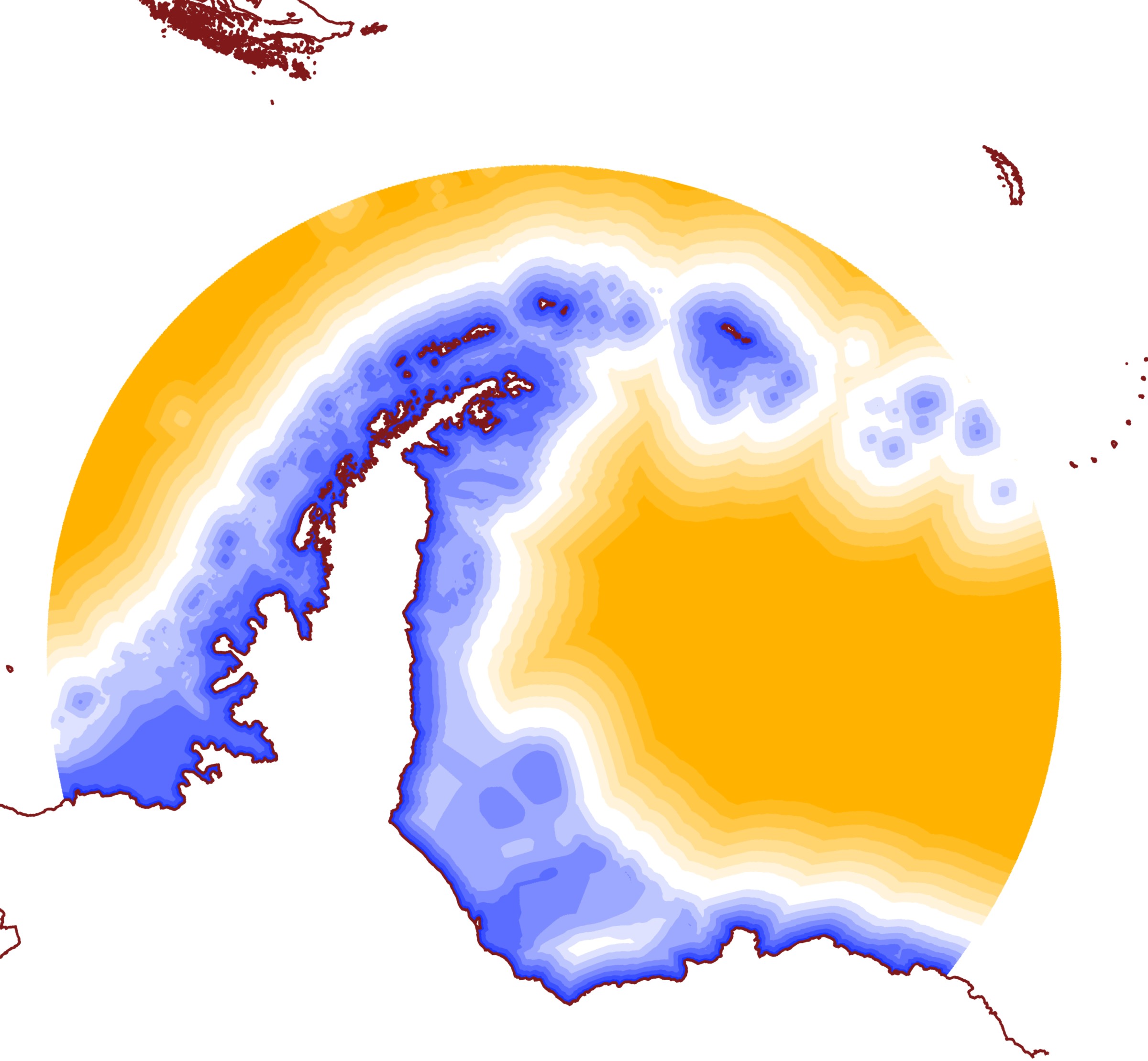}
\\[1ex]
\includegraphics[width=.9125\textwidth]{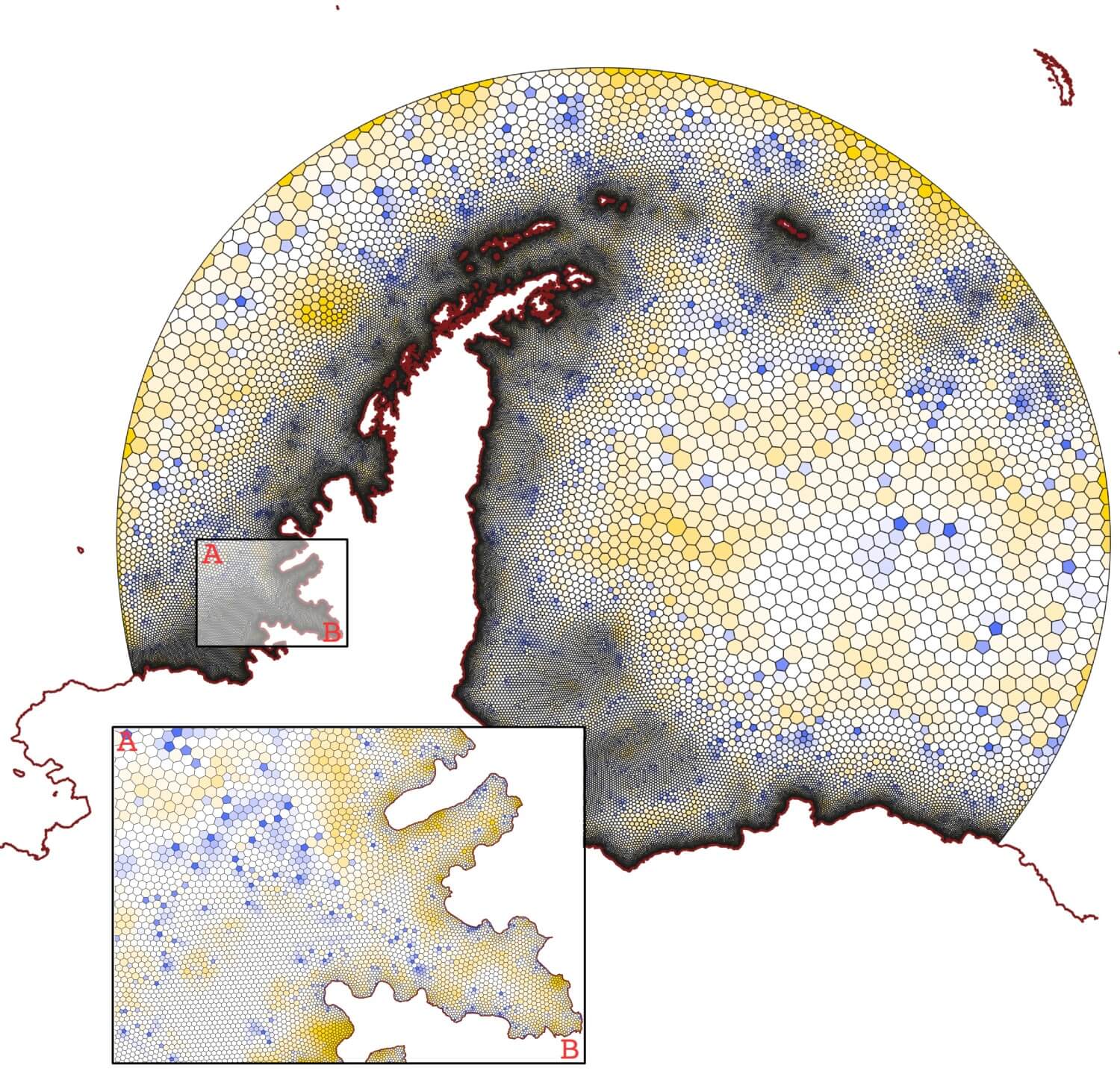}

\caption{An optimised primal-dual grid for the Antarctic test-case, showing (clockwise from top-left): (a) the coastal geometry and open-boundary definition, (b) the weighted triangulation $\mathcal{T}(\mathbf{X},W)$ (c) contours of the mesh spacing function $\bar{h}(\mathbf{x})$, and (d) the dual power diagram $\mathcal{D}(\mathbf{X},W)$. Here, dual cells are coloured by their \textit{relative power}, with blue tones indicating negative values, orange tones positive values, and white tones associated with values approaching zero.}

\label{fig_antarctica_mesh}
\end{figure*}

In the `Australian' and `Antarctic' test-cases shown in Figures~\ref{fig_australia_mesh} and \ref{fig_antarctica_mesh}, the performance of the coupled primal-dual optimisation algorithm developed in Section~\ref{section_primal_dual_optim} is investigated; seeking to generate optimal primal-dual pairs appropriate for a co-volume type discretisation scheme. These test problems were designed to explore the effectiveness of the proposed methods for the generation of very high-quality staggered unstructured grids for regional ocean modelling applications, focusing on the East Australian Shelf \& Coral Sea region in the `Australian' example, and the Antarctic Peninsula \& Weddell Sea region in the `Antarctic' problem. In both cases, a two-dimensional, coastally conforming geometry was generated by employing a locally-centred \textit{stereographic projection}; a conformal Spherical-to-Cartesian mapping designed to minimise distortion about a region of interest. Additional bathymetric data was retrieved from the well-known ETOPO1 dataset \cite{amante2009etopo1}; providing a high-resolution representation of the local sea-floor. These test-cases will form a basis for future modelling experiments using the Model for Predication Across Scales (MPAS-O) \cite{ringler2008multiresolution,ringler2013multi}, in which an unstructured ocean circulation model will be employed to study ocean dynamics in a multi-resolution framework; transitioning from global to coastal scales.  

In both problems, the Frontal-Delaunay refinement algorithm provided by the JIGSAW package \cite{engwirda2017jigsaw,JIGSAW-GEO} was used to produce an initial configuration, consisting of a high-quality, graded Delaunay triangulation and its associated Voronoi dual. The Frontal-Delaunay algorithm is known to produce particularly high-quality Delaunay triangulations \cite{engwirda2016off}; guaranteeing that all angles in the primal triangles exceed $30^\circ$, and that the distribution of element sizes is a tight approximation to the target grid-spacing function $\bar{h}(\mathbf{x})$. Use of the high-quality Delaunay-Voronoi grid generated by the Frontal-Delaunay refinement procedure can be thought of as providing a `good' initial condition for the subsequent primal-dual optimisation phase. {\reviewerB In this study, grid-spacing was scaled based on the maximum expected barotropic wave speed
\begin{gather}
\tilde{h}(\mathbf{x}) = \max \left( \tilde{h}_{\text{min}},\, \min \left(\tilde{h}_{\text{max}},\, \beta \sqrt{g D(\mathbf{x})} \right) \right) \,, 
\end{gather}
where $D(\mathbf{x})$ is the mean ocean depth, $g$ is the acceleration due to gravity and $\beta$ is a user-defined parameter used to control effective mesh resolution. The scalars $\tilde{h}_{\text{min}}$ and $\tilde{h}_{\text{max}}$ are lower and upper bounds, here set to $\tilde{h}_{\text{min}} = 5\mathrm{km}$ and $\tilde{h}_{\text{max}} = 50\mathrm{km}$. Such grid-spacing is designed to achieve a quasi-uniform CFL restriction on model time-step throughout the domain and to provide high mesh resolution in shallow regions on the shelf and in the coastal zone. Noting that bathymetry, and hence ocean depth $D(\mathbf{x})$ is typically non-smooth, a additional \textit{gradient-limiting} process is applied to the raw grid spacing function $\tilde{h}(\mathbf{x})$ to generate a Lipschitz-continuous grid-spacing function $\bar{h}(\mathbf{x})$, subject to
\begin{gather}
|\nabla \bar{h}(\mathbf{x})| \leq g\,, \quad \text{where} \quad g \in \mathbb{R}^{+}\,.
\end{gather}
Following the work of Persson \cite{Persson06SizeFunc}, the smooth function $\bar{h}(\mathbf{x})$ is taken as the steady-state solution of the associated Hamilton-Jacobi equation
\begin{gather}
\partial_{t}\bar{h}(t,\mathbf{x}) + |\nabla \bar{h}(t,\mathbf{x})| = \min \left( g, |\nabla \bar{h}(t,\mathbf{x})| \right)\,, \quad \text{with} \quad \bar{h}(0,\mathbf{x}) = \tilde{h}(\mathbf{x})\,.
\end{gather} 
In this study, a scalar smoothing parameter $g \in \mathbb{R}^{+}$ is used to globally, and isotropically limit spatial gradients. The gradient-limited size function $\bar{h}(\mathbf{x})$ becomes more uniform as $g \rightarrow 0$. Here, $g = 0.1$ is used throughout.
}

\subsubsection{Discussions}

\medskip

\begin{figure*}
\centering
\rotatebox{90}{\qquad\quad \text{No optimisation}}\quad
\includegraphics[width=.3875\textwidth]{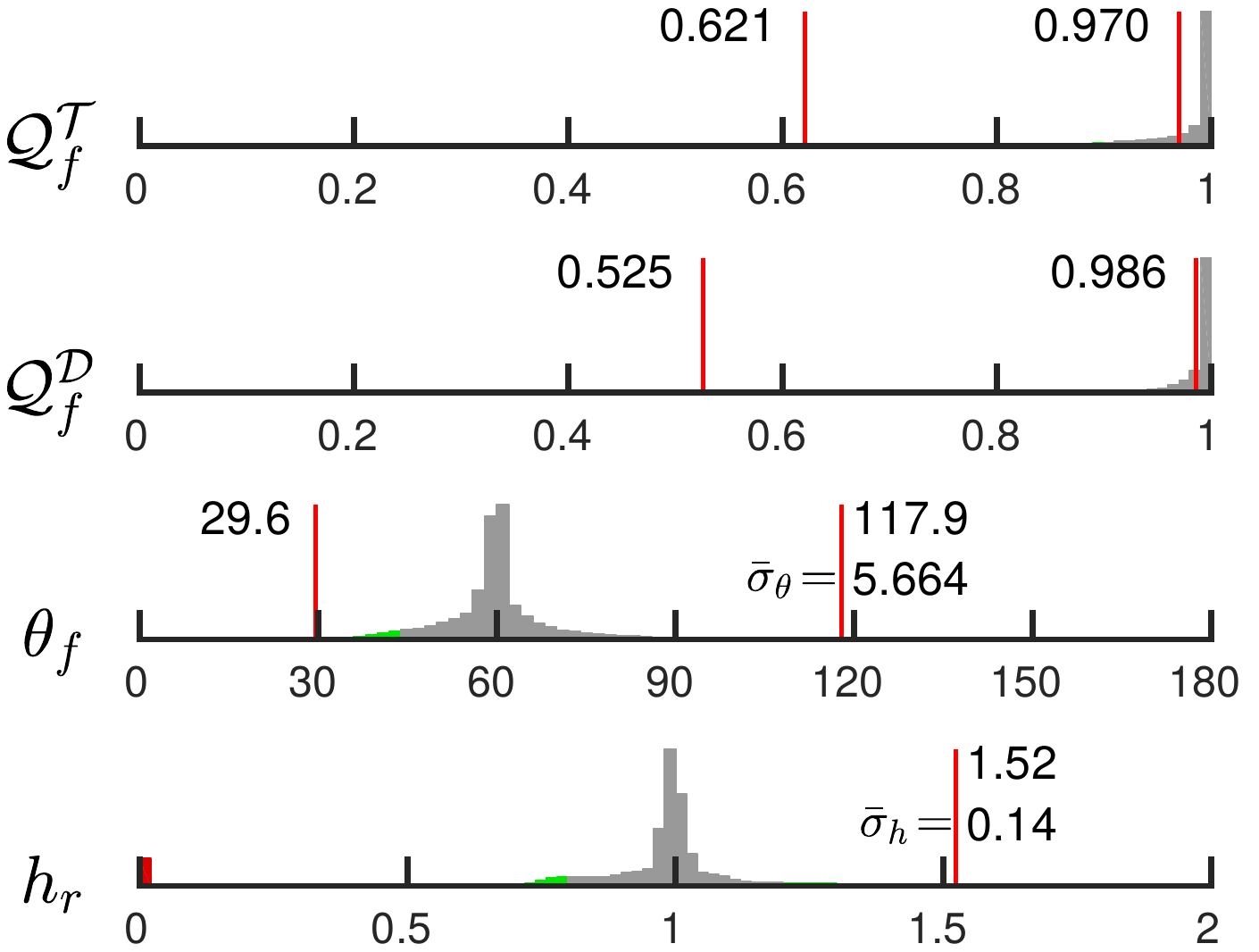}\qquad
\includegraphics[width=.3875\textwidth]{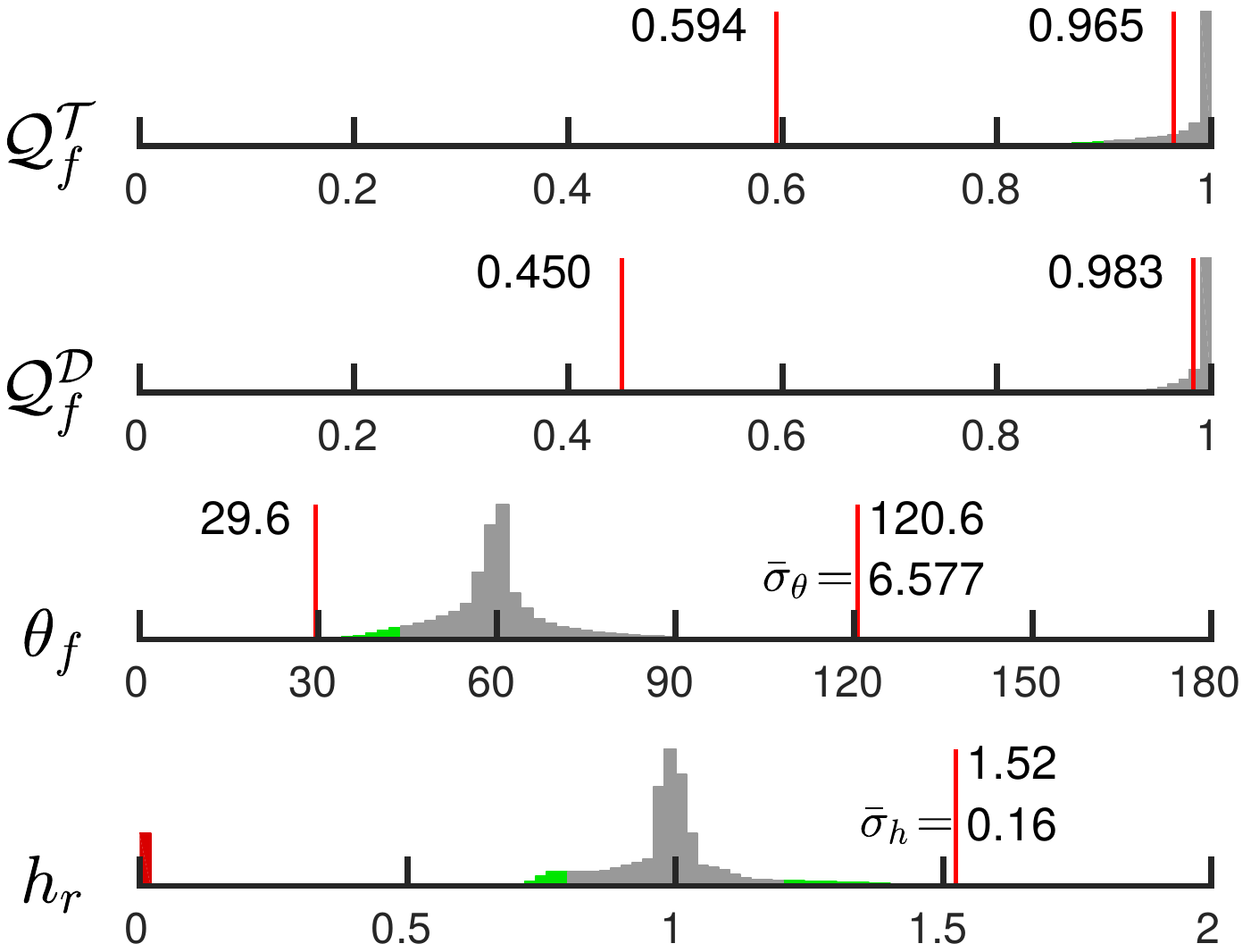}
\\[5ex]
\rotatebox{90}{\qquad\qquad \text{Primal-only}}\quad
\includegraphics[width=.3875\textwidth]{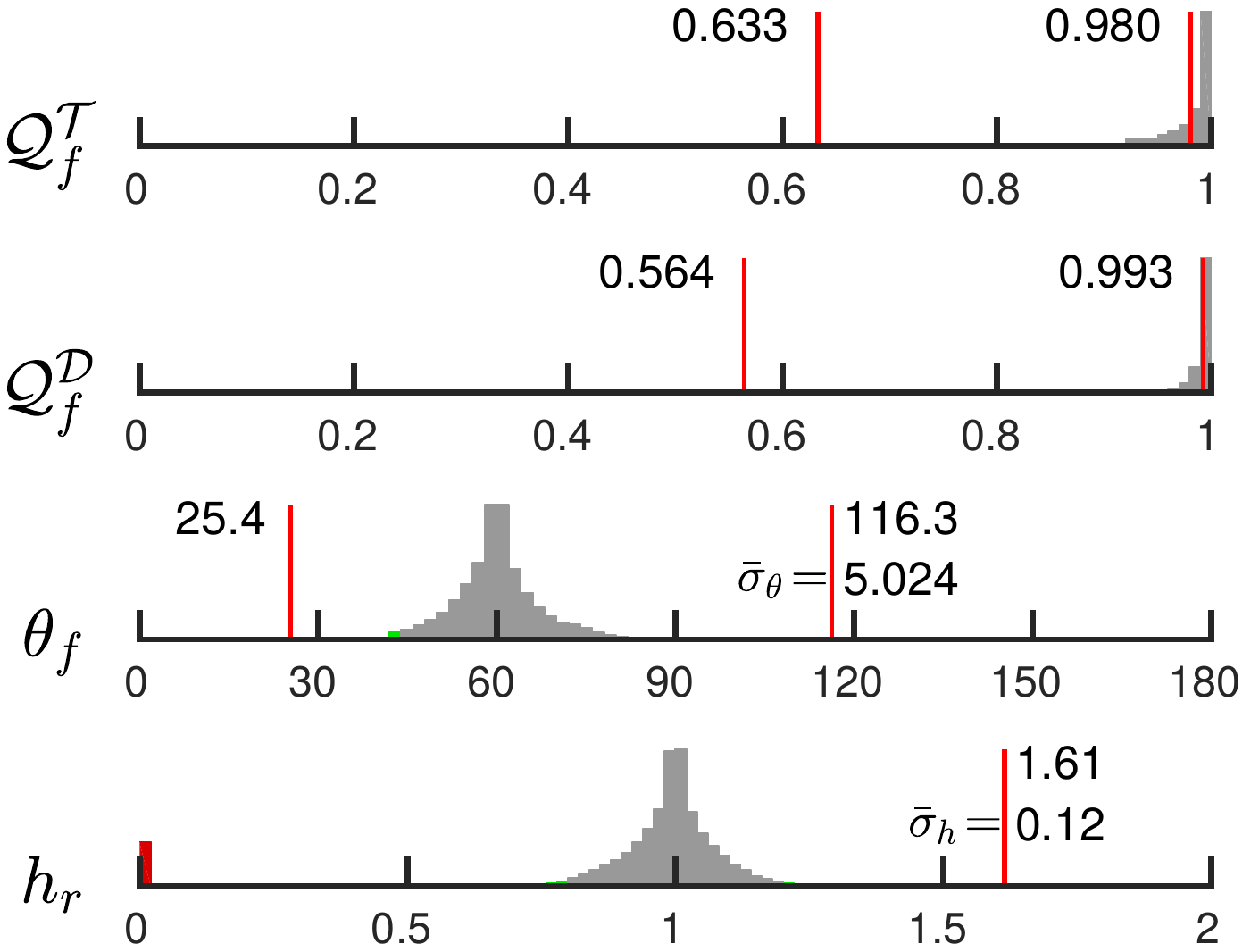}\qquad
\includegraphics[width=.3875\textwidth]{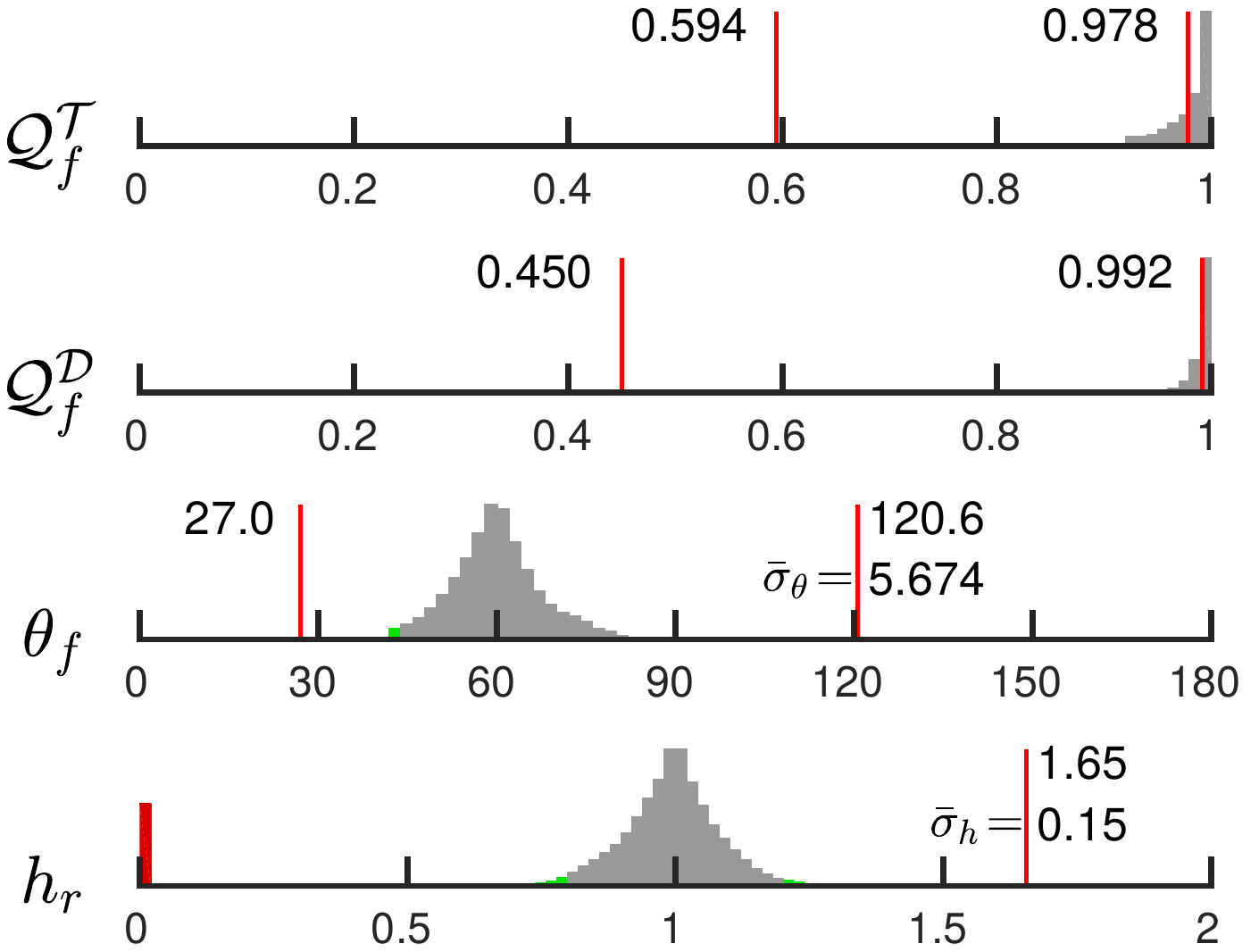}
\\[5ex]
\rotatebox{90}{\qquad\qquad \text{Primal-dual}}\quad
\includegraphics[width=.3875\textwidth]{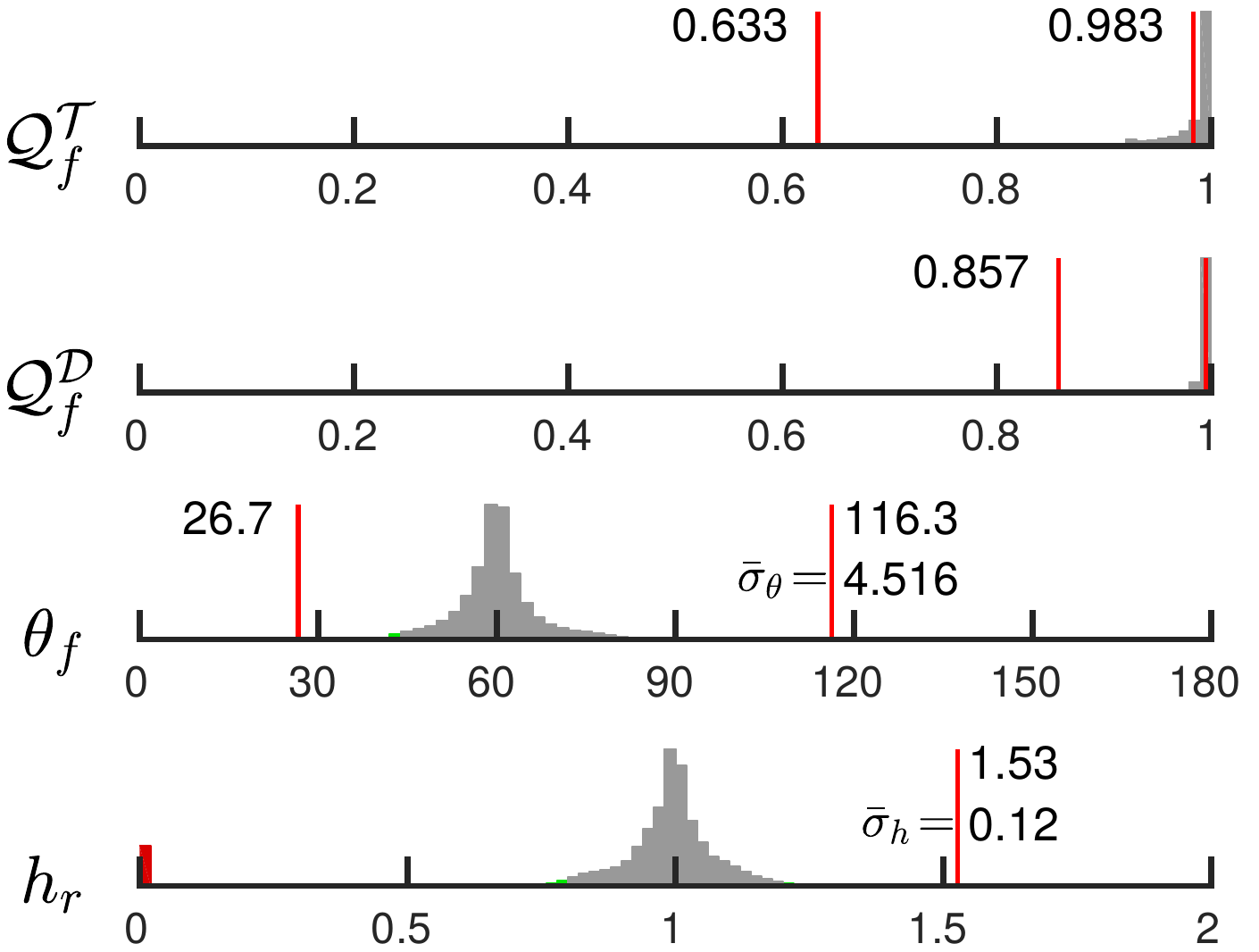}\qquad
\includegraphics[width=.3875\textwidth]{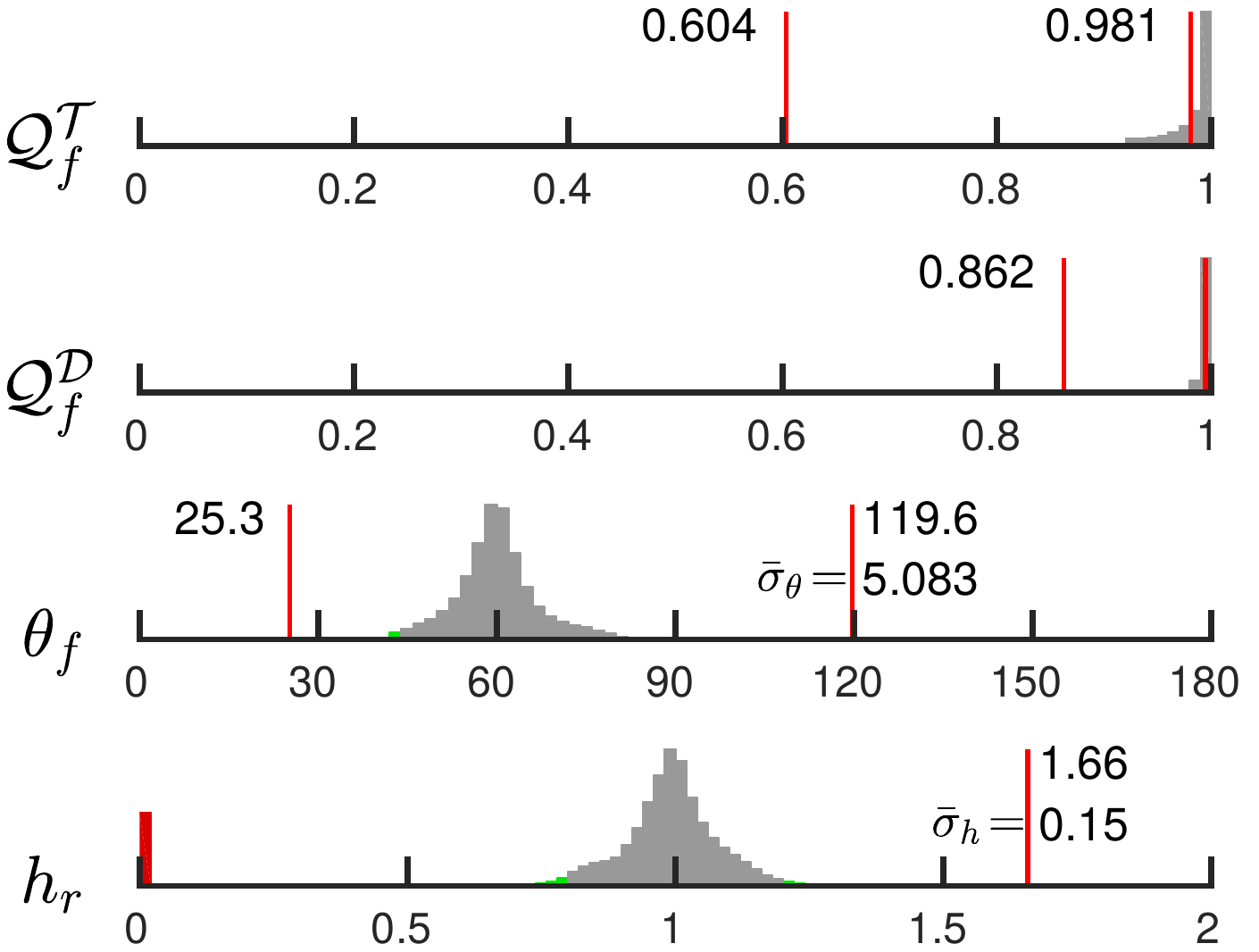}

\caption{Histograms of primal-dual quality metrics for the Australian (left) and Antarctic (right) test problems, showing metrics for (from top to bottom): (a) the initial grid, (b) application of the primal-only optimisation scheme, and (c) application of the new coupled primal-dual approach.}

\label{fig_grids_cost}
\end{figure*}

A set of optimised primal-dual pairs are presented in Figures~\ref{fig_australia_mesh} and \ref{fig_antarctica_mesh}, showing: (a) the geometry of each test problem, {\reviewerB (b) the primal triangulation $\mathcal{T}(\mathbf{X},W)$, (c) contours of the target grid-spacing function $\bar{h}(\mathbf{x})$,} and (d) detail of the associated dual structure $\mathcal{D}(\mathbf{X},W)$. Results are shown after the application of the primal-dual optimisation procedure. Dual cells are coloured by their relative power $W_{r}$, showing the distribution of weights throughout the grid. Mesh quality is quantified in Figure~\ref{fig_grids_cost}, where histograms of primal and dual grid quality metrics illustrate the distribution of $\mathcal{Q}^{\mathcal{D}}(\mathbf{X},W)$, $\mathcal{Q}^{\mathcal{T}}(\mathbf{X})$, $\theta(\tau)$ and $h_{r}(e)$ before and after the application of mesh optimisation. Results for three different configurations are presented: (i) the initial configuration, (ii) the result of primal-only optimisation, and (iii) the result of the new coupled primal-dual optimisation method presented in Section~\ref{section_primal_dual_optim}. The primal-only scheme employed in (ii) is based on an optimisation of the primal grid only, leading to the generation of standard Delaunay-Voronoi pairs with $W^{n} = \{0\}$. This primal-only scheme contains the full set of operations descried in Section~\ref{section_primal_dual_optim}: the hybrid ODT-type vertex smoothing operator, topological flips, and edge refinement/collapse operations, but does not include the weight-selection strategy described in Section~\ref{section_dual_optim}. This primal-only procedure is similar to the strategy presented by the author in \cite{engwirda2017jigsaw} for the optimisation of Delaunay-Voronoi pairs, and is used here to quantify the added benefit derived from a coupled optimisation of the full primal-dual grid structure compared to that of the primal tessellation alone.

The primal-dual pairs presented in Figures~\ref{fig_australia_mesh} and \ref{fig_antarctica_mesh} show that the coupled optimisation scheme is effective in generating high-quality, graded primal-dual tessellations for problems involving complex boundary definitions, such as coastlines, and non-uniform mesh spacing constraints. Visually, grid quality appears to be very high, with the dual grids consisting of a majority of regular hexagonal polygons, smoothly transitioning between regions of high and low resolution. The effectiveness of the coupled scheme is confirmed through an analysis of the grid quality metrics presented in Figure~\ref{fig_grids_cost}, showing that optimisation improves the minimum and mean values for both the primal and dual metrics $\mathcal{Q}^{\mathcal{T}}(\mathbf{X},W)$ and $\mathcal{Q}^{\mathcal{D}}(\mathbf{X},W)$, as well as sharpening the distribution of angles in the primal mesh about $\theta(\tau) \rightarrow 60^\circ$ and mesh-spacing conformance such that $h_{r} \rightarrow 1$. In the `Australian' test problem, application of the primal-dual scheme improves grid quality such that $\mathcal{Q}_{\text{min}}^{D}: 0.525 \rightarrow 0.857$, $\bar{\mathcal{Q}}^{D}: 0.986 \rightarrow 0.997$, $\mathcal{Q}_{\text{min}}^{T}: 0.621 \rightarrow 0.633$, and $\bar{\mathcal{Q}}^{T}: 0.970 \rightarrow 0.983$. Similar improvement was reported for the `Antarctic' test-case, with $\mathcal{Q}_{\text{min}}^{D}: 0.450 \rightarrow 0.862$, $\bar{\mathcal{Q}}^{D}: 0.983 \rightarrow 0.997$, $\mathcal{Q}_{\text{min}}^{T}: 0.594 \rightarrow 0.604$, and $\bar{\mathcal{Q}}^{T}: 0.965 \rightarrow 0.981$. The full set of results are summarised in Table~\ref{table_summary}.

Compared to the primal-only scheme, application of the coupled optimisation strategy was found to induce larger improvements in the primal and dual grid quality metrics for both test problems; demonstrating the effectiveness of such an approach. Interestingly, as well as significant improvements to the dual metrics $\mathcal{Q}^{D}(\mathbf{X},W)$, the coupled algorithm was able to improve the primal triangulations slightly more than the primal-only scheme, resulting in sharper distributions of $\theta(\tau)$ and slightly improved $\bar{\mathcal{Q}}^{\mathcal{T}}$. Such results present interesting questions for the development of ODT/CVT schemes, showing that, despite such schemes being derived using the standard Delaunay-Voronoi pair, they may benefit in practice from use of the generalised, weighted structures employed here. The coupled optimisation strategy was also found to be significantly more effective than the primal-only scheme in enhancing the well-centredness of the grids, reducing the number of poorly staggered triangles by a factor of 25-30 in both cases. {\reviewerC The small number of poorly-staggered elements not removed via optimisation were observed to span highly non-convex segments of the boundary, with all triangle vertices fixed on the bounding contour. In the present implementation, such configurations offer few opportunities for optimisation, admitting perturbations to weight values only. It is expected that these cases may be remediated by employing more sophisticated updates for boundary constrained entities, for example, allowing mesh vertices to `slide' along the bounding contour.} In practice, the generation of complex, well-centred Delaunay-Voronoi grids is known to be extremely difficult \cite{vanderzee2008well,vanderzee2010well}, showcasing the utility of the new Regular-Power structures presented here. Overall, these results demonstrate that the coupled, primal-dual optimisation scheme can be used to generate very high-quality staggered, pair-wise orthogonal unstructured grids that are both more centroidal and better self-centred than standard Delaunay-Voronoi type approaches.

Turning, lastly, to the distribution of vertex weights generated by the coupled scheme, the relative power metric $W_{r}$ is used to colour the cells of the dual grids presented in Figures~\ref{fig_australia_mesh} and \ref{fig_antarctica_mesh}. These results show that the distribution of weights throughout each mesh is typically smooth, except in regions where the spacing of the grid is changed abruptly, or in the neighbourhood of topological inconsistencies. It appears that dual cells with a lower than optimal valence (i.e.~5-cells and smaller) are typically associated with larger, negative weights. Conversely, those with higher than optimal valence (i.e.~7-cells and larger) are typically associated with larger positive weights. Larger magnitude weights were also observed in the neighbourhood of boundary constraints, where it is often more difficult to optimise the geometry and/or topology of the grid via other means. Perfectly regular hexagonal configurations were found to induce more uniform distributions of weights; typically tending towards zero. Such behaviour is consistent with expectations; showing that local weight gradients are induced in patches of the grid where a conventional Delaunay-Voronoi pair would define sub-optimal configurations.

\section{Conclusions \& future work}
\label{section_conclusions}

A new optimisation-based algorithm designed to generate very high-quality staggered unstructured grids has been described, focusing on the development of techniques for the construction of weighted Regular-Power tessellations appropriate for unstructured co-volume type numerical formulations. Employing a combination of geometrical, topological and weight-based local optimisation operators, the new coupled primal-dual mesh optimisation schemes have been shown to generate generalised staggered unstructured grids that are simultaneously pair-wise orthogonal, centroidal and are typically well-centred. The performance of the new schemes have been tested experimentally, demonstrating that improved primal-dual complexes are generated compared to conventional optimised Delaunay-Voronoi type tessellations. Specifically, results confirm that 
the generalised Regular-Power pairs generated by the new optimisation algorithms are typically both more centroidal and are better self-centred. 


A range of work is slated for future investigations, focusing on further improvements to the grid generation algorithm itself, as well as an application of the new methodology to various problems in ocean modelling and computational physics more broadly. Work on the primal-dual optimisation algorithm is proceeding in a number of directions, including: (i) an investigation of alternative, quasi-Newton based optimisation schemes \cite{chen2011efficient,chen2014revisiting}; exploring whether further enhancements to grid quality can be achieved through the application of globally-coupled optimisation procedures, (ii) the generation of higher-dimensional primal-dual tessellations through an extension of the Regular-Power formulation to problems in $\mathbb{R}^{3}$, and (iii) improvements to the computational efficiency of the implementation, through the application of parallel programming patterns. Such work will be made available through future releases of the JIGSAW meshing library; currently under development by the author \cite{JIGSAW-GEO}. 

{\reviewerC
It is expected that the methods described here can be generalised to higher-dimensional problems, both for triangulations embedded in $\mathbb{R}^{d}$ and for complexes of higher topological degree (i.e.~tetrahedral meshes). In the case of general $d$-simplex meshes, additional work is needed | firstly, requiring the definition of an appropriate dual quality metric and secondly, 
necessitating the development of methods to update general $d$-simplex topologies. It is proposed that a straightforward extension of the dual metric (\ref{eqn_dual_quality}) described here could be derived for higher degree elements, consisting of a weighted sum of the geometrical `defects' associated with each $(d-k)$-face per simplex. For example, a dual quality metric for tetrahedral complexes would consist of a weighted sum of the six edge-, four face- and single element-centred defect terms associated with each tetrahedron. It is proposed that optimisation of such a metric would reduce the geometrical offsets between the various $(d-k)$-faces in a given primal-dual pair, leading to configurations that are more mutually `centred'.

It is expected that topological updates may be more challenging for higher-degree problems. Specifically, it is known \cite{ChengDeyShewchuk} that algorithms based on local element flips are not guaranteed to recover the topologies of higher-dimensional Delaunay and/or Regular tessellations; potentially becoming trapped in local topological optima. It is proposed that a full re-triangulation would instead be computed at each iteration, based on, for example, efficient implementations of Bowyer-Watson type tessellation schemes \cite{Boissonnat02CGALTria}. While such strategies may increase computational expense compared to the flipping-based scheme described here, they are not expected to fundamentally alter the composition of the resulting primal-dual optimisation scheme, which would consist of optimisation-driven vertex and weight updates, a global re-triangulaton step, and various edge- and face- collapse and refinement operations.  
}

The use of generalised primal-dual structures as a basis for computational simulation is a relatively new area of interest, and a wide range of investigations are expected to be fruitful. It is expected that the application of optimal tessellations that are more centroidal and better self-centred than conventional Delaunay-Voronoi pairs may lead to improved numerical performance; enhancing the accuracy, stability and efficiency of the underlying formulations. Presently, the generation of optimal unstructured pairs for problems in geophysical fluid dynamics is a key focus; seeking to optimise the performance of the mimetic co-volume scheme employed in the Model for Prediction Across Scales (MPAS-O) \cite{ringler2008multiresolution,ringler2010unified,ringler2013multi} for the simulation of ocean dynamics and climate processes. The application of generalised grids to other problems in computational physics is also of interest, including, for example, applicability to co-volume type formulations for the solution of Maxwell's equations (e.g.~\cite{yee1966numerical,walton2017advances}), in addition to various methods formulated using Discrete Exterior Calculus.

\section{Acknowledgements} 

This work was conducted at the Centre for Climate Systems Research (CCSR) at Columbia University and NASA's Goddard Institute for Space Studies. The author wishes to thank Scott Mitchell for his thoughts on an initial draft, as well as the three anonymous reviewers for their helpful comments and feedback.

\section{Code availability}

The JIGSAW mesh generation package used in this study is available online: \url{https://github.com/dengwirda/jigsaw-geo-matlab}.

\section{References}

\bibliographystyle{elsarticle-num}
\bibliography{references}

\end{document}